\documentclass[5p,preprint,sort]{elsarticle}

\usepackage[flushleft]{threeparttable}
\usepackage{lipsum} 
\usepackage{multirow}
\usepackage{array}
\usepackage{url}
\usepackage{subcaption}
\usepackage{dirtytalk} 
\usepackage{pgf-pie} 

\definecolor{color1}{RGB}{212,209,218} 
\definecolor{color2}{RGB}{244,214,198} 
\definecolor{color3}{RGB}{251,243,209} 
\definecolor{color4}{RGB}{228,233,215}
\definecolor{color5}{RGB}{201,229,255} 
\definecolor{color6}{RGB}{233,222,211} 
\definecolor{color7}{RGB}{255,231,231} 
\definecolor{color8}{RGB}{250,221,202} 
\definecolor{color9}{RGB}{207,212,194} 
\definecolor{color10}{RGB}{243,190,190}
\definecolor{color11}{RGB}{253,200,220}
\definecolor{color12}{RGB}{235,180,230}

\newcolumntype{L}[1]{>{\raggedright\let\newline\\\arraybackslash\hspace{0pt}}m{#1}}
\newcolumntype{C}[1]{>{\centering\let\newline\\\arraybackslash\hspace{0pt}}m{#1}}
\newcolumntype{R}[1]{>{\raggedleft\let\newline\\\arraybackslash\hspace{0pt}}m{#1}}

\newtheorem{defn}{Definition}

\usepackage{soul}
\sethlcolor{white}

\begin{document}\sloppy
	
\begin{frontmatter}
		
\title{XAI-CF -- Examining the Role of Explainable Artificial Intelligence in Cyber Forensics}
		
\author[sa]{Shahid Alam\corref{cor1}}
\ead{sha.alam@uoh.edu.sa}

\author[za]{Zeynep Altiparmak}
\ead{zzeynepaltiparmak@gmail.com}
		
\address[sa]{Department of Information Security, College of Computer Science and Engineering, University of Ha’il, Ha’il, Saudi Arabia}
\address[za]{Department of Computer Engineering, Adana Alparslan Turkes Science and Technology University, Adana, Turkey\\\tt{The article has been accepted for publication in Engineering Applications of Artificial Intelligence\\The final published version is available at https://doi.org/10.1016/j.engappai.2026.113892}}
		
\cortext[cor1]{Principal corresponding author, sha.alam@uoh.edu.sa}
	
\begin{abstract}
\hl{With the rise of complex cyber devices, Cyber Forensics (CF) is facing numerous new challenges. For instance, many systems are operating on smartphones, each hosting millions of downloadable applications. Analyzing this vast amount of data and making sense of it requires innovative techniques, particularly those from Artificial Intelligence (AI). To successfully implement these techniques in CF, it is essential to justify and explain the results to CF stakeholders. This will enable informed decision-making and foster trust in AI systems. An explainable AI (XAI) system can fulfill this role in CF, and we refer to this system as XAI-CF. Although XAI-CF is crucial, it is still in its early stages of development. Thus, there is a need to investigate and advocate for the importance and benefits of XAI-CF. Furthermore, previous studies do not provide a comprehensive survey of XAI's role in CF. Many only briefly address the challenges and ways to enhance XAI for this field, and none have proposed a unified framework for an XAI-CF system. To address these research gaps, this paper discusses the key requirements and prerequisites for an XAI-CF system. We present a thorough literature review of prior works that apply and utilize XAI to build and enhance trust in CF. In addition to discussing the challenges faced by XAI-CF and offering concrete solutions, we introduce the first XAI-CF framework that cohesively integrates XAI principles across every stage of the CF lifecycle. We believe that our work provides a solid foundation for future researchers interested in XAI-CF.}

\end{abstract}


\end{frontmatter}

\section{Introduction and Motivation}

Forensic analysis is a major part of investigating a security incident and involves detecting, collecting, and documenting the pieces of evidence. Most of the time, forensic analysis is linked with evidence for the court of law. When these pieces of evidence are found on a cyber device, then their analysis is called Cyber Forensics (CF). One of the first documented and practical examples of CF investigation was carried out in 1986~\citep{stoll1988stalking, stoll2005cuckoo}. The intruder was caught by tracing the network activities and performing a forensics analysis of different computer logs. At that time, forensics was performed by computer professionals on an ad-hoc case-by-case basis. The forensics analyst in the case \citep{stoll1988stalking} was an Astronomer~\citep{stoll2005cuckoo}.

With the rise of complex cyber devices such as smartphones, Internet of Things, and control systems in automotive and drones, the proliferation of operating systems and file formats, pervasive encryption, use of the cloud for remote processing and storage, and legal standards, CF is facing many new challenges. For example, there are dozens of systems running on smartphones, each with more than a million downloadable applications. Sifting through this large amount of data and making sense requires new techniques, such as those from the field of Artificial Intelligence (AI). A detailed list and discussion of challenges in CF can be found in \citep{garfinkel2010digital, du2020sok}.

The first documented research proposal on AI \citep{AI-modern-approach-2020} was initiated by McCarthy et al. \citep{mccarthy2006proposal} in 1955. The major goals discussed in the proposal were: automation; how to program a computer to use a language; neuron nets for forming concepts; and self-improvement. Some of the current traditional goals of AI research include reasoning; decision-making; knowledge representation; natural language processing; and Machine Learning (ML). Some of the major techniques that AI uses to achieve these goals are statistics; artificial neural networks; formal logic; cognitive psychology; and linguistics. Most of the recent breakthroughs in modern AI came through deep learning \citep{AI-modern-approach-2020}. AI is a powerful technology and has potential benefits, but also comes with potential risks and challenges, such as lack of transparency, inherently unexplainable, especially when using deep learning, adversarial attacks, privacy, misinformation, algorithm and data bias, and ethics. For example, the recently developed ChatGPT \citep{ChatGPT}, the new AI language model, has revolutionized the approach in AI and is being used in various domains, such as business, education, healthcare, etc. The potentials of ChatGPT are unlimited, but it can be a double-edged sword \citep{fui2023generative}. For example, for businesses, it can create creative content such as ideas for advertisements, and on the other hand, it can create hallucinations and produce fake information. Another example is the use of ChatGPT in higher education which can promote collaboration among students and remote learning, and can also be used for improving student assessments, such as grading essays, assignments, exams, and quizzes more quickly and accurately \citep{cotton2023chatting}. However, using ChatGPT for assessment comes with some challenges, such as the possibility of plagiarism \citep{cotton2023chatting}. ChatGPT can generate essays based on a set of parameters or prompts, which means students can generate and submit essays that are not their own work \citep{dehouche2021plagiarism}. It becomes very difficult to distinguish between a student's own writing and the responses generated by ChatGPT. Moreover, the responses generated by ChatGPT may not accurately reflect the true level of understanding of the student. For a detailed list of risks and challenges, interested readers are referred to \citep{scherer2015regulating,AI-modern-approach-2020,fui2023generative,ray2023chatgpt}.

AI has been successfully applied in various fields as well as in Cyber Security, but its application and utilization specifically in CF is still relatively limited \citep{automation-ai-df-2021}. To bridge this gap, we need to examine and explore the potential effects of AI in CF and develop standardized techniques for mining cyber (digital) evidence \citep{df-ai-2022}.	Due to the complexities involved in the process of CF, which requires an intelligent analysis sifting through and making sense of different patterns, the use of AI in this field is particularly valuable \citep{ai-df-2014}. The lack of transparency of an AI system gives rise to different challenges in the field of CF. One of the main challenges is that it requires validation by a human user \citep{hall2022explainable}. This opacity makes it difficult to admit a CF analysis using AI in a court of law because it is often taken as a black box that is impossible to justify or explain up to legal standards \citep{ai-df-2014}.

One of the most important factors in accepting the use of AI in CF is developing trust in AI systems. Some of the other factors in accepting the use of AI in CF are to make AI authentic, interpretable, understandable, and interactive. This way, AI systems will be more acceptable to the public and ensure alignment with legal standards. An explainable AI (XAI) system can play this role in CF, and we call such a system XAI-CF. XAI adds transparency and accountability to an AI system and helps create trustworthy AI. This helps a human investigator to understand and validate AI systems/algorithms, which makes them more reliable and trustworthy \citep{xai-cybersecurity-survey-2022}. XAI rationalizes the workings and justifies the outputs of an AI system \citep{hall2022explainable}. XAI-CF is indispensable and is still in its infancy. In this paper, we explore and make a case for the significance and advantages of XAI-CF. We strongly emphasize the need to build a successful and practical XAI-CF system and discuss some of the main requirements and prerequisites of such a system.

There are different stakeholders of XAI \citep{preece2018stakeholders}. Each one of them has different requirements for XAI. \textit{Developers} of AI applications want XAI to aid in testing, debugging, and evaluation. \textit{Theorists} who are advancing AI theory want to understand XAI to advance the state of the art in XAI rather than develop practical XAI applications. \textit{Ethicists} who are policy makers, and critics of AI, are concerned about the accountability and transparency of XAI. \textit{Users} of XAI are the most influential stakeholders and contribute the most to growing the literature on XAI. In terms of the CF, they are part of the users' stakeholders. Members of the CF community need explanations to better understand the output of an AI system, decide how to act on the given outputs, and justify their actions. These actions play an important role in making key and accountable decisions in a court of law.

\begin{figure*}[htb]
	\centering
	\includegraphics[scale=0.72]{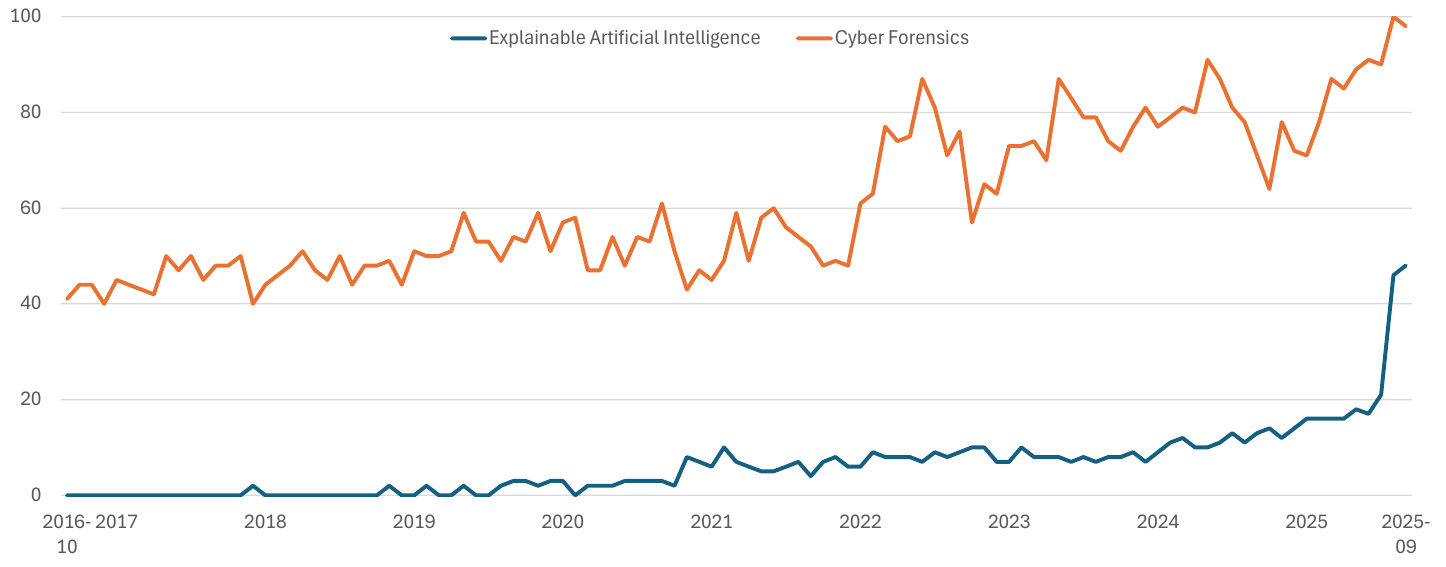}
	\caption{Google search trends \citep{Google-search-trends} by month for the last ten years, for the terms \textit{Explainable Artificial Intelligence} and \textit{Cyber Forensics}. The y-axis depicts the relative search frequency for the term. A value of 100 is the peak popularity for the term. A value of 50 means that the term is half as popular. A score of 0 means there was not enough data for this term.}
	\label{fig:Google-trends-XAI-CF}
\end{figure*}

The graphs shown in Figure \ref{fig:Google-trends-XAI-CF} depict the popularity by month of the two terms \textit{Cyber Forensics} and \textit{Explainable Artificial Intelligence} when used during Google searches performed in the last ten years (2016 -- 2025). This shows the development in the use of terminology based on the current most popular search engine, which is useful and of value to identify trends. Starting from the year 2016 the term \textit{Cyber Forensics} remains popular and even gains in popularity after the year 2022. There are a few mentions of the term \textit{Explainable Artificial Intelligence} in the years from 2016 -- 2021. After 2021, the term XAI started gaining popularity. In 2025, XAI has seen a substantial rise in popularity, although its search volume is still roughly 50\% lower than that of CF. Still, XAI is not considered a popular term in comparison to CF, and there is a need to develop awareness in the general public of its importance and relevance in specific domains. This is one of the motivations of this paper to make a case for the significance and advantages of XAI in CF and strongly recommend its use in CF. Based on the popularity results from this graph, the study in this paper reviews the works on XAI, CF, and XAI-CF only for the last ten years (2016 -- 2025).

\subsection{Contributions}

\hl{Despite the rising interest in XAI from 2021 to 2025, no survey has focused solely on its application in CF. Existing works do not provide a comprehensive review of XAI’s role in cyber forensics, and only a handful briefly mention the challenges and potential improvements for applying XAI in this area. In this paper, we examine and justify the importance and benefits of XAI-CF.

This work concentrates specifically on XAI in the context of CF. Readers seeking broader discussions on the role and impact of AI in CF may consult}~\citep{iqbal2020artificial, vasilaras2024artificial, chaudhary2025artificial}. The main contributions of this paper are as follows:

\begin{itemize}
	\item
	In this paper, we explore and make a case for the significance and advantages of XAI-CF. We strongly emphasize the need to build a successful and practical XAI-CF system and discuss some of the main requirements and prerequisites of such a system.
	\item
	We present a formal definition of the terms CF and XAI-CF and a comprehensive literature review of the previous works carried out that apply and utilize XAI to build and increase trust in CF.
	\item
	To make the reader familiar with the study carried out in this paper, in addition to the background, we also present a critical and short review of the works carried out in the last ten years in XAI and CF.
	\item
	We discuss some challenges facing XAI-CF, such as adversarial attacks, bias management, oversimplification, the CF and AI chasm, and human-computer interaction, and also provide some concrete solutions to these challenges.
	\item
	\hl{The paper presents the first XAI-CF framework that systematically integrates XAI principles throughout all phases of the cyber forensics lifecycle.}
	\item
    We identify key insights and future research directions for building XAI applications for CF. This paper is an effort to explore and familiarize the readers with the role of XAI applications in CF, and we believe that our work provides a promising basis for future researchers interested in XAI-CF.
\end{itemize}

\subsection{Definitions}\label{sec:defs}

Before defining CF, we present standard definitions of the terms Cyber and Forensics by the Cambridge Dictionary.
\textit{Cyber} \citep{def-cyber} -- Computers, especially the Internet, and activities that use them.
\textit{Forensics} \citep{def-forensics} -- Scientific methods of solving crimes, that involve examining objects or substances related to a crime. Based on these definitions, we define:

\begin{defn}
	\textbf{Cyber Device} is a piece of electronic hardware that can process and/or store data and can connect to the Internet. These include a server, desktop computers, laptops, smartphones, IoT, and embedded systems, such as a router, modems, network adapters, storage devices, control systems in automotive and drones, industrial control systems, etc.
\end{defn}

\begin{defn}\label{def:CF}
	\textbf{Cyber Forensics (CF)} uses scientific methods and expertise for gathering and analyzing pieces of evidence found in cyber devices that can be used in criminal or other investigations in a court of law. This evidence can be used for different purposes, such as electronic discovery, intelligence, and administration.
\end{defn}

Before defining XAI-CF in this paper, we present the definition of XAI from other sources. The first definition presented by David Gunning \citep{gunning2017explainable} states \say{The Explainable AI (XAI) program aims to create a suite of ML techniques that enable human users to understand, appropriately trust, and effectively manage the emerging generation of artificially intelligent partners.} Another definition is by Arrieta et al. \citep{arrieta2020explainable} that states \say{Given an audience, an XAI system produces details or reasons to make its functioning clear or easy to understand.} The term \textit{Explainable} means able to explain, and the Cambridge Dictionary defines the term \textit{Explain} \citep{def-explain} -- To make something clear or easy to understand by describing or giving information about it. Based on all these definitions, we define:

\begin{defn}\label{def:XAI-CF}
	\textbf{XAI-CF} is a set of techniques that describes and provides authentic information about an AI system employed in CF in such a way that stakeholders of CF can understand and trust the AI system. These techniques can be scientific or non-scientific or a combination of both depending on the level of comprehension and expertise of the stakeholders of CF in the domain (type of knowledge) of the AI system.
\end{defn}

Four main groups form the major \textbf{\textit{Stakeholders of CF}}: (1) developers of CF; (2) CF analysts; (3) members of the court, such as the judges, lawyers, and jury members; (4) and anyone legally involved in the case, such as the accused, perpetrators, and victims. We have included the fourth group as stakeholders because they are also interested in learning how the AI system works. This can help them carry out adversarial attacks using anti-forensics techniques [76] to evade detection by the XAI system during forensics analysis.

\subsection{Organization of the Paper}

\begin{figure*}[!ht]
	\centering
	\includegraphics[scale=0.54]{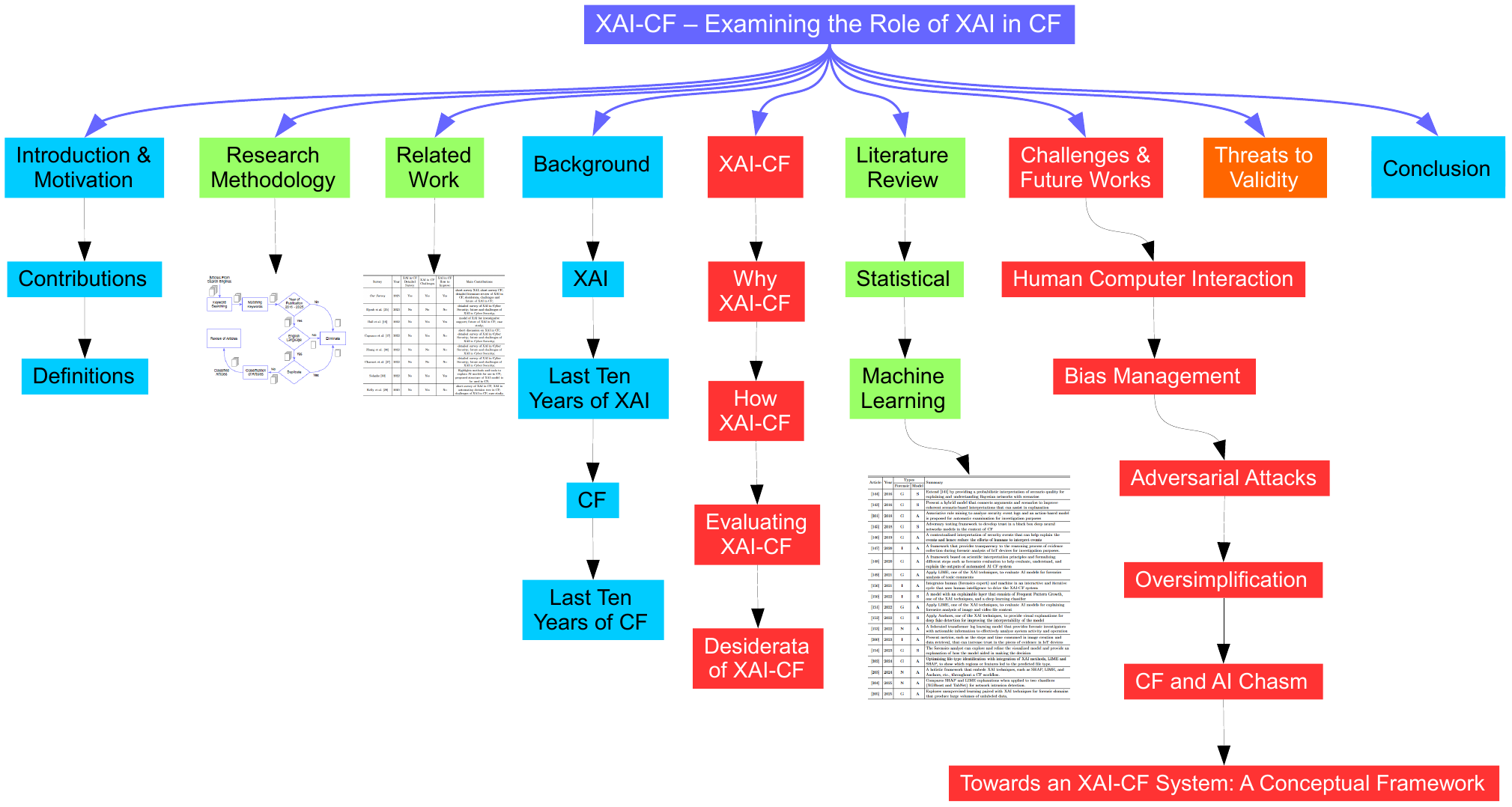}
	\caption{\hl{Overview of the paper’s structure and progression.}}
	\label{fig:paper_structure}
\end{figure*}

The rest of the paper is organized as follows. Section \ref{sec:rm} describes the methods used to carry out the survey in this paper. Section \ref{sec:related-work} details the related works (works similar to our work), i.e., the survey/review of the works about the application of XAI in CF. Section \ref{sec:background} gives background information about XAI and CF and also presents a short critical review of the last ten years of XAI and CF. Section \ref{sec:XAI-CF} discusses why, how, evaluation, and desiderata of XAI-CF. Section \ref{sec:litreview} presents a comprehensive literature review of previous works carried out in the area of XAI-CF. Section \ref{sec:challenges} discusses the challenges of building a practical XAI-CF system/framework and provides solutions to these challenges as part of future work. Section \ref{sec:threatstovalidity} discusses the threats to the validity of the study presented in this paper. Finally, section \ref{sec:conclusion} concludes our work. Figure~\ref{fig:paper_structure} provides an overview of the paper’s structure and progression.

\section{Research Methodology}\label{sec:rm}

\begin{figure*}[!ht]
	\centering
	\includegraphics[scale=0.55]{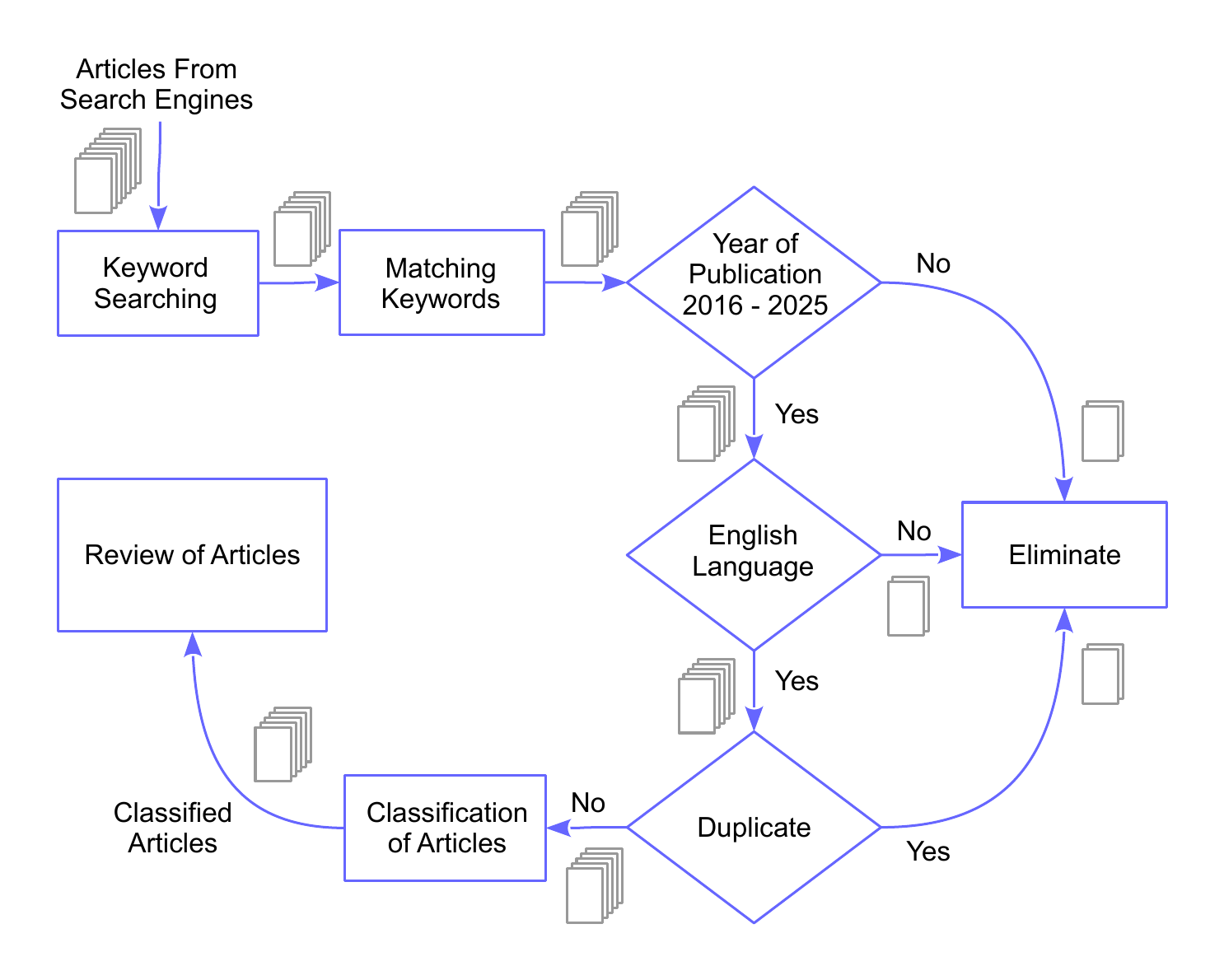}
	\caption{\hl{Flowchart of the research methodology/process used for the survey carried out in this paper.}}
	\label{fig:rm_flowchart}
\end{figure*}

The goal of this work is to examine and investigate the research carried out in the area of XAI-CF. Figure \ref{fig:rm_flowchart} shows the flowchart of the research methodology used for the survey carried out in this paper. The flowchart is developed and used to carry out the \textit{four surveys/reviews} in this paper. Two short reviews, one on the \textit{Last Ten Years of XAI} in section \ref{sec:last-10-XAI} and the second on the \textit{Last Ten Years of CF} in section \ref{sec:last-10-CF}. Two comprehensive reviews, one is the \textit{Related Work} in section \ref{sec:related-work} and the second is the \textit{Literature Review} in section \ref{sec:litreview}. To increase the credibility of the study carried out in this paper, we used the principle of \textit{triangulation} \citep{yin2015qualitative,patton2014qualitative}, i.e., seeking at least three ways of verifying a procedure, piece of data, or finding.

The initial step in the flowchart involves the collection of articles from various \textbf{Search Engines}. This ensures a comprehensive and diverse set of literature related to the study carried out in this paper. 
Having gathered a pool of articles, the next step is to employ a \textbf{Keyword Searching} mechanism. Keywords serve as the navigational tools in this vast sea of information, helping to filter and identify articles that specifically address the intersection of XAI and CF. This step is crucial for narrowing down the scope of the literature review and extracting articles that directly contribute to understanding the role of XAI in the context of CF. The selected keywords were: \textit{Artificial Intelligence}, \textit{Explainable Artificial Intelligence}, \textit{Cyber Forensics}, \textit{Digital Forensics}, and \textit{Challenges}. We used either a single or a combination of these keywords. We also searched for the relevant papers referenced in the articles found through keyword searching.

As part of the \textit{data triangulation}, we used \textit{scite} (\url{https:\\www.scite.ai} -- an AI-powered research engine), \textit{Google Scholar search}, and \textit{manual confirmation} of the collected data (articles) from different sources. By initiating the research methodology with this systematic approach, the aim is to ensure that subsequent phases are built upon a solid foundation of relevant and focused academic literature. 

With the collected articles and the foundation laid by the keyword search, the next step in the flowchart is the \textbf{Matching Keywords} step. In this step, the goal is to refine the selection further by evaluating each article for the presence of specific keywords. This filtering process enhances the precision and specificity of the article selection, aligning with the research focus of the paper. There were several keywords used during this matching process. In addition to the above keywords, some of the other keywords used in this process were \textit{Interpretable}, \textit{Understandable}, \textit{Interactive}, \textit{Authentic}, \textit{Visualization}, and \textit{Adversarial Attacks}. 

\begin{figure*}[!ht]
	\centering
	\begin{subfigure}{.48\textwidth}
		\begin{tikzpicture}[scale=1]
			\pie[rotate=0,color={color1, color2, color3, color4, color5, color6, color7, color8, color9}]{
            27.5/Elsevier, 
            27.5/IEEE,     
            10.2/Springer, 
            8.7/MDPI,     
            5.8/ArXiv,    
            5.8/Nature,   
            14.5/Others    
			}
		\end{tikzpicture}
	\end{subfigure}
	\begin{subfigure}{.48\textwidth}
		\centering
		\begin{tikzpicture}[scale=1]
			\pie[rotate=0,color={color1, color2, color3, color4, color5, color6, color7, color8, color9,color10}]{
            7.2/2016, 
            4.4/2017, 
            5.8/2018, 
            5.8/2019, 
            17.4/2020, 
            7.2/2021, 
            20.3/2022, 
            11.6/2023, 
            10.2/2024, 
            10.2/2025 
			}
		\end{tikzpicture}
	\end{subfigure}
	\caption{\hl{Distribution of the 69 selected articles for classification and review by source and year. The source \textit{Others} include one paper (1.5\%) each from ACM, OUP, Wiley, AAAI, PeerJ, NDSS, Hindawi, IntechOpen, PLOS, and ISMIR.}}
	\label{fig:sources_years}
\end{figure*}
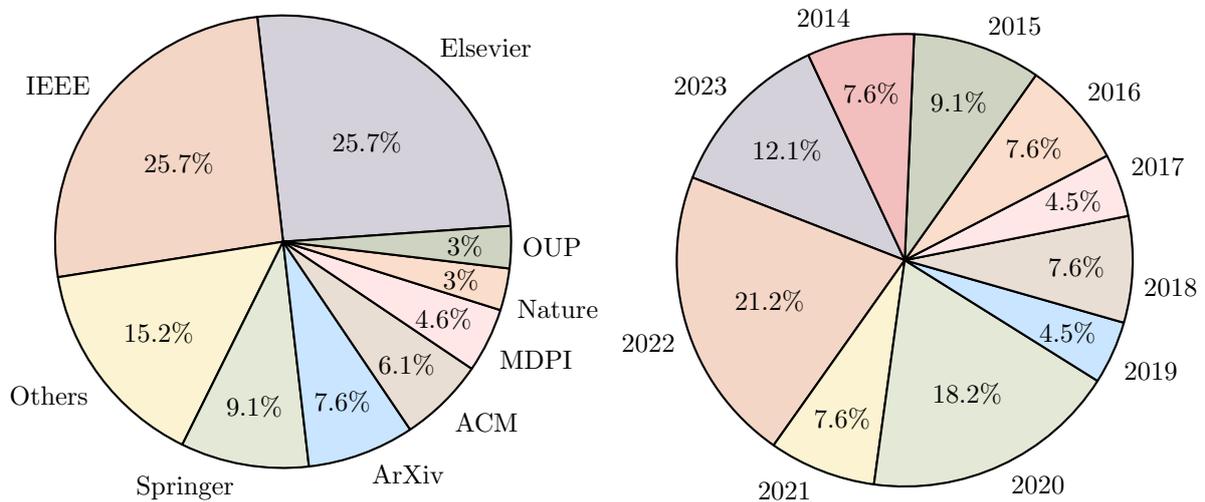

\hl{Then, we examine whether the \textbf{Publication Year} of each selected article falls within the range of 2016 to 2025.}
If the answer is yes, indicating the relevance and recency of the literature, the flowchart proceeds to the next step, where the articles undergo an \textit{English language} check. However, if the answer is no, the flowchart directs the articles to the \textit{Eliminate} stage, as they may not align with the desired time frame for the research. This step ensures that the chosen articles are not only pertinent to the research focus but also recent enough to reflect the current state of XAI in the context of CF. 
After the English language comes the \textbf{Duplicate Check} stage. If the article is duplicated, it is directed to the elimination stage. 
\hl{After this stage, the total number of articles selected for classification and review was 69.}
Distribution of these selected articles by source and year is shown in Figure \ref{fig:sources_years}. These articles are reviewed in sections \ref{sec:related-work} , \ref{sec:last-10-XAI} , \ref{sec:last-10-CF} , and \ref{sec:litreview} in the paper.

Having ensured the uniqueness of the selected articles and duplicates elimination, the flowchart proceeds to the \textbf{Classification of Articles} step. In this step, the articles are systematically grouped based on their primary focus. The reason for classifying articles into groups is to provide a structured approach to analyzing the role and application of XAI in CF. We classify the articles in the literature review, section \ref{sec:litreview}, carried out in this paper into two categories.

\textit{Statistical Category:} Articles falling under this classification predominantly emphasize statistical methods, approaches, and analyses within the context of XAI in CF. The focus here is on articles that contribute to the statistical foundations of interpretability and transparency in CF.

\textit{Machine Learning Category:} This classification encompasses articles that primarily delve into ML aspects concerning XAI in CF. These may include studies on algorithmic transparency, model interpretability, and the application of ML techniques in CF scenarios.

With the articles successfully classified into distinct categories the next pivotal step is the \textbf{Review of Articles}. This phase involves a detailed examination of each group, aiming to extract valuable insights, patterns, and findings. The detailed review of classified articles serves as a critical phase for synthesizing findings, identifying gaps in the existing literature, and gaining a deeper understanding of the nuances within each thematic category. 

As part of the \textit{methodological triangulation}, we used three reviewers to check and confirm the procedures and methods adapted, as depicted in the flowchart, for the study carried out in this paper.

In summation, this methodical journey through the research methodology, from collecting articles and refining them through keyword searches to the meticulous classification and review stages, has provided a comprehensive understanding of the role of the studies carried out in this paper. By carefully curating and categorizing articles into different domains, this approach ensures a nuanced exploration of interpretability and transparency in the context of the study. The insights gleaned from this systematic process contribute to a more profound comprehension of the subject, paving the way for meaningful discussions, insightful implications, and potential advancements in the realm of XAI applied to CF.

\section{Related Work}\label{sec:related-work}

There are several works \citep{adadi2018peeking, arrieta2020explainable, javed2023survey} that have reviewed/surveyed the research works carried out on the application of XAI in different domains and fields. In this section, we discuss the previous works that have specifically explored (surveyed/reviewed) in some context the application and utilization of XAI in CF.

The work carried out in \citep{kelly2020explainable} studies the challenges of developing XAI systems to automate the decision-making process in the generation of evidence in CF. The challenges discussed, such as a large number of features, missing data, etc., are mostly related to XAI systems. However, one of the challenges of how to drive a decision tree automatically using the available forensic evidence data is specifically related to CF. There is no comprehensive survey or review of previous and current works presented in the paper about the application and utilization of XAI in CF. The case study of drug testing presented in the paper collects medical forensics lab data, such as hair and nails, through a structured questionnaire, and then digitizes it for forensic analysis.

The work carried out in \citep{solanke2022explainable} proposes recommendations for an interpretable AI-based CF. Some of the recommendations for mitigating distrust in AI discussed in the paper are: as far as possible, use less complex, non-opaque techniques; the methods should be evaluated based on their ability to articulate the explanation; and the incorporation of a human-in-the-loop so that validations can be performed before reaching a decision. The paper also highlights some of the methods for explaining AI models, such as model simplification, feature importance, visualization etc. Like the previous work \citep{kelly2020explainable} this paper highlights the shortcomings and challenges of AI and how they can be improved for successfully applying AI in digital forensics. There is no comprehensive survey on the application of XAI in CF.

Recently, several works \citep{reynaud2025review, mohale2025systematic, sharma2025comprehensive, manthena2025explainable, rjoub2023survey,xai-cybersecurity-survey-2022,zhang2022explainable,charmet2022explainable} present a comprehensive survey on the application and utilization of XAI in Cyber Security. The main focus of the works is on Cyber Security, and the only work that briefly discusses CF is \citep{xai-cybersecurity-survey-2022}. All of these works present a detailed literature review of the works that seek to achieve explainability in the following sub-fields of Cyber Security: Network security, Endpoint security; Application security, Intrusion detection systems, Malware detection, Phishing and Spam detection, and BotNet detection. \citep{xai-cybersecurity-survey-2022} presents a short discussion about fraud detection, zero-day vulnerabilities, digital forensics, cyber-physical systems, and crypto-jacking.

The work carried out in \citep{hall2022explainable} is the closest to our work. They also explore the potential of XAI to enhance CF. The main focus of this study is to present a model of XAI that can be applied to a CF case for investigative support. The authors also present a fictional case study of XAI for investigative support. The paper also discusses the future of AI as XAI in CF and how it is going to change the field of CF. However, there is no comprehensive survey or review of previous and current works presented in the paper about the application and utilization of XAI in CF.

Table \ref{tab:comparison} presents a comparison of previous surveys with the survey carried out in this paper. As we can see, unlike other works, the work carried out in this paper covers all the major aspects of a successful and practical XAI-CF system. If we want to apply AI successfully in CF, then we need to develop trust in AI systems by making them authentic, interpretable, understandable, and interactive. The role of XAI in CF is indispensable, and we need more research work in this area. This is the main motivation of this paper to explore the role and application of XAI-CF. Based on our extensive search and the discussion provided above, this is the first study to deliver a comprehensive survey and review of XAI-CF applications and usage, while also proposing the first unified XAI-CF framework that spans all stages of the cyber forensics process.

\begin{table*}[!ht]
	\caption{\hl{Comparison of our survey/review with seven other similar works discussed in Section {\ref{sec:related-work}}}.}
	\setlength{\tabcolsep}{4pt}
	\renewcommand{\arraystretch}{1.3}
\begin{center}
	\begin{tabular}{ c | c | C{1.65cm} | C{1.65cm} | C{1.65cm} | C{1.75cm} | C{5.6cm} } \hline\hline
		Survey                 & Year & XAI in CF Detailed Survey & XAI in CF Challenges & XAI in CF How to Improve & \hl{Unified XAI-CF Framework} & Main Contributions \\ \hline
		\emph{Our Survey}	                                 & 2025 & Yes & Yes & Yes & Yes & short survey XAI; short survey CF; detailed literature review of XAI in CF; desiderata, challenges and future of XAI in CF; \\ \hline
		Rjoub et al. \citep{rjoub2023survey}                 & 2023 & No  & No  & No  & No & detailed survey of XAI in Cyber Security; future and challenges of XAI in Cyber Security; \\ \hline
		Hall et al. \citep{hall2022explainable}              & 2022 & No  & Yes & Yes & No & model of XAI for investigative support; future of XAI in CF; case study; \\ \hline
		Capuano et al. \citep{xai-cybersecurity-survey-2022} & 2022 & No  & Yes & No  & No & short discussion on XAI in CF; detailed survey of XAI in Cyber Security; future and challenges of XAI in Cyber Security; \\ \hline
		Zhang et al. \citep{zhang2022explainable}            & 2022 & No  & No  & No  & No & detailed survey of XAI in Cyber Security; future and challenges of XAI in Cyber Security; \\ \hline
		Charmet et al. \citep{charmet2022explainable}        & 2022 & No  & No  & No  & No & detailed survey of XAI in Cyber Security; future and challenges of XAI in Cyber Security; \\ \hline
		Solanke \citep{solanke2022explainable}               & 2022 & No  & Yes & Yes & No & highlights methods and tools to explain AI models for use in CF; proposed structure of XAI model to be used in CF; \\ \hline
		Kelly et al. \citep{kelly2020explainable}            & 2020 & No  & Yes & No  & No & short survey of XAI in CF; XAI in automating decision tree in CF; challenges of XAI in CF; case study; \\ \hline
	\end{tabular}
\end{center}
	\label{tab:comparison}
\end{table*}

\section{Background}\label{sec:background}

\subsection{Explainable Artificial Intelligence (XAI)}\label{sec:XAI}

A formal definition of XAI is given in section \ref{sec:defs}. This section explores in more detail of why, what, and how of XAI. The omnipresence of AI dictates how powerful it is and has become a part of our lives. It is being applied in making critical decisions, especially in the field of health \citep{jimma2023artificial,afnan2021interpretable} and forensic sciences \citep{galante2023applications}. There are certain issues facing AI when applied in these domains. One of the major issues is trust. For example, when a forensic analyst identifies the suspect of a child exploitation case with the help of the reports generated by an AI system. At the time this suspect is presented in a court of law with the evidence, will the court be able to give a judgment based on these findings? To trust these findings, the court will require an explanation of these findings, and the AI methods used to reach the findings will need to be interpreted and explained. This is one of the core requirements of any forensic analysis during a cybercrime investigation to be accepted and presented in a court of law. The main purpose of XAI is to develop trust in AI and play a central role in a successful legal CF analysis.

There are three basic approaches to generating explanations: (1) Using simple models, such as decision trees and other rule-based approaches. The output of these models depends on a small set of rules that are easy to explain to the system's user. (2) Feature selection and analysis, focusing on a small set of features. This makes it easy to understand the relationship between input and output, even if the model is complex. (3) The most prevalent approach uses external mechanisms to help describe the inner workings of a black-box AI system. One of the methods in this approach creates a proxy model, such as a decision tree, to understand and then explain the original model. One other method perturbs the model by querying the model with different inputs to assess which inputs are important. Popular XAI algorithms within this approach are LIME (Local Interpretable Model-Agnostic Explanations) \citep{ribeiro2016should} and SHAP (SHapley Additive exPlanation) \citep{lundberg2020local,lundberg2017unified}. The National Institute of Standards and Technology introduced four principles \citep{phillips2021four} and proposed that all XAI systems should adhere to these principles.

\begin{enumerate}
	\item
	\textit{Explanation} -- A system is explainable if its outcome can be reasoned logically, intelligently, and methodically. Every AI system in practice has a different way of explaining itself. Therefore, to keep it generic, this principle does not put any condition of correctness or impose any quality metrics. There are two types of XAI models, self-explainable and post-hoc explainable. \textit{Self-Explainable} as the name suggests, explain themselves. The model can be explained globally as well as locally. Each decision on an input can be explained, also called local explanation, and an explanation can also be provided for all the inputs as a whole, also called a global explanation. Most of the common self-explanatory models are decision trees and regression. These are some of the simplest AI models and are designed to be more meaningful to humans. \textit{Post-Hoc Explainable} as the name suggests, explains after the output is seen. These explanations are rendered by other software tools and can be used on the algorithms without knowing their inner workings, provided the outputs can be queried over the inputs. These explanations produce feature importance scores, rule sets, heatmaps, or natural language representations. This helps a user quantify the importance of features and explain the decisions produced by the model. Some researchers \citep{caruana2020intelligible,dovsilovic2018explainable,linardatos2020explainable} argue the trade-off of simplicity with accuracy and state that a self-explainable model is less accurate than a post-hoc model. There are other researchers \citep{rudin2019stop,rudin2019we,afnan2021interpretable} that disagree with this statement.
	\item
	\textit{Meaningful} -- An AI system is meaningful if its explanation is understandable. Stakeholders of the AI system play an important role in this principle. Whether an explanation is meaningful depends on the stakeholders' knowledge, experiences, relationship to the system, and various other psychological factors. The other major factor is the purpose of the explanation, such as it will be different for legal forensics than for a business purpose. This makes the explanation intelligible to the intended audience. There are some challenges in making the explanation meaningful for a variety of different audiences. For example, a forensic scientist explains the evidence to a layperson, e.g., a member of a jury. Such an explanation in general is difficult to understand and may mislead a layman reader, and therefore does not fulfill the criteria for being meaningful \citep{jackson2015communicating}. Some of the challenges \citep{jackson2015communicating,lombrozo2006structure,mueller2019explanation} for producing a meaningful explanation are individual differences of the users about a meaningful explanation, the context of an explanation, and the expectation of users about different explanation types. An XAI system should provide meaningful information appropriate to the context and understandable by the intended users.
	\item
	\textit{Explanation Accuracy} -- This is the ability of the explanation to accurately describe how the AI system reaches its conclusion. This is different than the accuracy of an AI algorithm and system. The level of detail in the explanation is also important. For some audiences, simple explanations suffice; others may require more details to better understand the outcome. An inventor may explain her/his invention with substantial technical details to colleagues. The same explanation will likely be presented differently to untrained friends and parents. The accuracy of explanations depends on how meaningful they are to certain audiences. Therefore, meaningfulness and explanation accuracy are both associated. There are metrics for measuring the accuracy of an AI system \citep{japkowicz2011evaluating,phillips2000feret,phillips2007frvt,phillips2008iris,przybocki2007nist,reynolds20172016,hossin2015review}, such as receiver operating characteristic curve and area under the curve \citep{Introduction-ROC-Analysis}, precision-recall curve \citep{Precision-Recall}, and cost curves \citep{Cost-Curves} etc. Besides these, there is also work \citep{hoffman2018metrics,rosenfeld2021better,sovrano2023objective,lopes2022xai} in progress on developing metrics for evaluating the performance of an XAI system. How to measure the accuracy of an XAI system? There are certain criteria, such as, how good (precise and clear) the explanation is, how satisfied it is for the respective audience, and how trustworthy (justifiable) the explanation is. A detailed taxonomy of the XAI system's evaluation methods is given in \citep{lopes2022xai}.
	\item
	\textit{Knowledge Limits} -- This principle makes sure that the AI system operates within the limits of the system's knowledge and design. Any operation or outcome that is out of these limits produces a false judgment and hence is not trusted. This helps the system to dismiss misleading and ambiguous outputs/judgments and hence increases the trust in an AI system. There are two ways an AI system can exceed its knowledge limits. One is if the query or operation is outside the domain of the AI system. As an example, in CF, successful file fragment classification becomes challenging for an AI algorithm when an image is embedded in a file that itself is not an image, e.g., an image embedded in a PDF file \citep{alam2023sift}. In this case, if the AI algorithm classifies the fragment as an image rather than a PDF file without any explanation, then the system using the AI algorithm will not be able to dismiss this as a misleading output. But if the AI algorithm knows its limitations and explains then the AI system will be able to dismiss this misleading output. The second is if the confidence of the output is too low. Revisiting the above example, classifying a fragment in a multi-class (several file types) is also a challenging problem \citep{alam2023sift}. In this case, if the AI algorithm classifies a fragment into more than one class and randomly chooses one class out of these as an output, then this will be a misleading output. An AI system with the knowledge of these limitations will not trust such output.
\end{enumerate}

\subsubsection{Classification of XAI Methods}
\begin{figure}[!ht]
	\centering
	\includegraphics[scale=0.5]{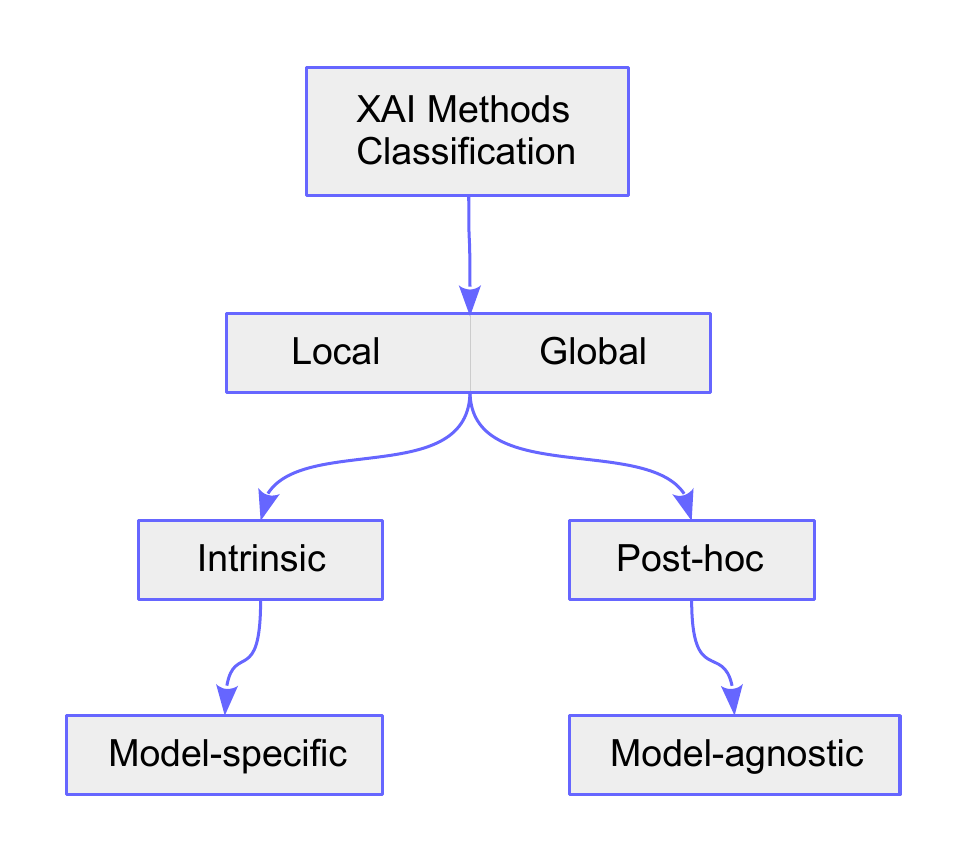}
	\caption{Classification of XAI Methods.}
	\label{fig:XAI_CF-methods}
\end{figure}

We divide and classify the XAI methods by the scope of their explanation as shown in Figure \ref{fig:XAI_CF-methods}. The scope of an explanation refers to its range and extent. It can be either local or global. Both local and global can be intrinsic and post-hoc. By definition, intrinsic methods are model-specific, and post-hoc methods are generally model-agnostic. This section presents some of the popular local and global methods. For a comprehensive coverage on this specific subject, readers are referred to \citep{vilone2020explainable,guidotti2018survey,arrieta2020explainable}.

\begin{enumerate}
	\item
	\textit{Local Explanations} -- The scope of these explanations is a single input or a subset of inputs. The most common is the explanation for the output of a single input point. Some of the local explanation algorithms are LIME \citep{ribeiro2016should}, SHAP \citep{lundberg2020local,lundberg2017unified}, Saliency Map \citep{iyer2018transparency,selvaraju2017grad}, and Counterfactual \citep{wachter2017counterfactual}. \textit{LIME} explains each prediction of the original model by building a proxy self-interpretable model, such as linear models or decision trees. This new model is trained on a new dataset containing the interpretable representations of the original data. The original model can be explained by the weights given to features in the new model. For example, a model predicts that a patient has the flu. LIME in the explanation highlights the symptoms with relative weights, such as sneezing and headache are given more weight than no fatigue. A doctor can easily decide with the help of these explanations. \textit{SHAP} also builds and uses a simpler model that is the interpretable approximation of the original model. It provides feature importance (contribution of each feature to the prediction result) by computing the Shapley value for each feature based on game theory. \textit{Saliency Map}, also known as sensitivity map or heat map, explains each prediction by computing an attribute value for each feature based on its degree of influence on the prediction result. A saliency map is a visual explanation of a model where contributions can be visualized as heat maps. \textit{Counterfactual} explanations indicate what amount of change in the input can modify the output. The objective is to minimize the new input value and bring it as close to the original input as possible. The goal is to maximize this objective function by a local search. This function depends on the endpoint loss, i.e., how close its value is to the desired class and the distance between the origin and the new value. Counterfactual explanations are similar to adversarial examples, but they differ in purpose. Unlike adversarial examples, where the purpose is to evade a classifier, they facilitate understanding of a model by providing explanations.
	\item
	\textit{Global Explanations} -- The scope of these explanations is the entire algorithm. Some of the global explanation algorithms are PDP (Partial Dependence Plot) \citep{friedman2001greedy}, ICE (Individual Conditional Expectation) \citep{goldstein2015peeking}, GSA (Global Sensitivity Analysis) \citep{cortez2011opening,cortez2013using}, SP-LIME (Submodular Pick LIME) \citep{ribeiro2016should}. \textit{PDP} explanations indicate the marginal change of the output when there is a change in the feature value. The PDP focuses on the overall average instance rather than a specific one. It reveals the relationship between a feature and the prediction results. \textit{ICE} plots are the graphs with line charts that represent the functional relationship between a feature and the predicted response. All the other features are kept fixed, only changing the value of the feature under analysis. \textit{GSA} method perturbs the input features through their range of values and quantifies its effects on the predictions of a given model. \textit{SP-LIME}, a variant of LIME, chooses the most relevant local LIME explanations to summarize and provide a global explanation.
\end{enumerate}

\subsection{Last Ten Years of XAI}\label{sec:last-10-XAI}

To make readers familiar with the study carried out in the paper, we present a critical and short review of the last ten years of work carried out in XAI. We exclude papers that explore the application of XAI in CF because we discuss in detail such applications in section \ref{sec:litreview}. A detailed survey on this topic is out of the scope of this paper. For a detailed survey, interested readers are referred to \citep{adadi2018peeking,ali2023explainable}.


\hl{Hara et al. {\citep{hara2016making}} propose a post-processing method to enhance the interpretability of Additive Tree Models (ATMs) {\citep{cui2015optimal}} by approximating them with simpler, more human-interpretable models. The challenge addressed is the high number of regions that ATMs typically create, hindering interpretability.}
Mishra et al. \citep{mishra2017local} extend the \textit{LIME} technique to music content analysis (MCA) named SLIME, introducing three versions of explanations based on temporal, frequency, and time-frequency segmentation. SLIME provides insights into model behavior, revealing limitations in the decision tree model's generalization despite high accuracy and demonstrating model-agnostic explanations for the neural network, aligning with model-dependent saliency maps in many cases.
Fong et al. \citep{fong2017interpretable} propose a model-agnostic framework for learning explanations for any black box algorithm, with a focus on identifying the image regions most responsible for classifier decisions. The proposed method employs explicit and interpretable image perturbations, addressing the limitations of heuristic-based image saliency methods. The framework contributes to interpretable and testable explanations, revealing insights into neural network fragility and susceptibility to artifacts.
Ribeiro et al. \citep{ribeiro2018anchors} propose a model-agnostic explanation system called \textit{Anchors}, which provides high-precision rules in the form of if-then statements to elucidate the behavior of complex models. Anchors efficiently compute explanations for any black-box model with high probability, showcasing their flexibility across various domains and tasks. A user study demonstrates that Anchors enable users to predict model behavior on unseen instances with higher precision and less effort compared to existing model-agnostic explanation techniques.
Guidotti et al. \citep{guidotti2018local} introduce \textit{LORE} (for LOcal Rule-based Explanations), a model-agnostic method for explaining black box decisions, focusing on providing interpretable and faithful explanations for specific instances. LORE employs a genetic algorithm to generate a synthetic neighborhood and then builds a local interpretable predictor in the form of a decision tree. The resulting explanation includes a decision rule and a set of counterfactual rules. LORE outperforms existing methods in terms of explanation quality and accuracy in mimicking the black box, addressing the ethical issue of a lack of transparency in decision systems.

Joo et al. \citep{joo2019visualization} investigate the application of \textit{Grad-CAM} to Deep Reinforcement Learning (DRL) in the context of Atari games. By revealing the significance of the CNN layer, the research enhances our understanding of how AI agents make decisions while playing Atari games through the application of Grad-CAM to DRL.
Gramegna et al. \citep{gramegna2020buy} propose an XAI model for deciphering customer decisions in the non-life insurance domain. The authors integrate XGBoost with Shapley Values, which enhances behavioral segmentation, showcasing the potential of explainable machine learning to refine our understanding of customer behavior in the insurance industry.
Cavaliere et al. \citep{cavaliere2020parkinson} address the challenge of poor explainability in complex machine learning models, particularly in medical applications. Their research focused on automatic Parkinson's disease detection, utilizing \textit{Grammar Evolution} (GE) with Automatically Defined Functions.
Parsa et al. \citep{parsa2020toward} apply the eXtreme Gradient Boosting (XGBoost) machine learning technique to enhance traffic safety through rapid accident detection. The study demonstrated XGBoost's effectiveness with high accuracy and detection rates, emphasizing the significance of traffic-related features, especially speed variations, through \textit{SHAP} analysis.
Feichtner et al. \citep{feichtner2020understanding} aims to identify inconsistencies between developer-defined application behavior and permission usage in Android applications. It utilizes a deep neural network trained on descriptive texts to predict permission scores, with \textit{LIME} employed for explainability analysis.
De et al. \citep{de2020explainable} propose a method to enhance the interpretability of decisions made by Deep Neural Networks (DNNs) by combining \textit{Cluster-TREPAN}, a decision tree generation algorithm, with hidden-layer-clustering. The focus is on explaining predictions from neural networks, emphasizing the importance of interpretability and reliability. The proposed Cluster-TREPAN method outperforms LIME in performance, effectively capturing and explaining the information flow within DNNs, addressing a critical aspect for human acceptance of deep learning models.

Szczepański et al. \citep{szczepanski2021new} focused on enhancing the explainability of BERT-based fake news detectors, addressing the crucial need for understanding model decisions in real-life scenarios. Utilizing surrogate-type explainability methods, specifically \textit{LIME} and \textit{Anchors}, the study demonstrates the feasibility of enhancing interpretability in linguistics-based fake news detection models. 
Gite1 et al. \citep{gite2021explainable} utilize an advanced combination of efficient machine learning techniques, specifically Long Short Term Memory (LSTM) for stock price prediction. The paper introduces an XAI element with \textit{LIME} for transparent and meaningful insights into the model's decision-making process.
Naeem et al. \citep{naeem2022explainable} address the critical issue of automated malware detection in IoT devices, leveraging deep learning techniques, and using \textit{Grad-CAM} to understand and improve the overall performance.
Sutton et al. \citep{sutton2022artificial}  apply the \textit{Grad-CAM} technique to enhance the diagnostic precision of ulcerative colitis, utilizing 8000 endoscopic images. While the choice of Grad-CAM provides visual insights into model learning, a more detailed discussion of the AI techniques used and the specific XAI methods employed would provide a comprehensive evaluation of the study's contributions.
Debjit et al. \citep{debjit2022improved} work presents an advanced machine learning framework for early COVID-19 detection, leveraging the Harris Hawks Optimization algorithm to optimize ML classifiers' hyperparameters and \textit{SHAP} values for model interpretation. Utilizing SHAP, the research offers transparency by elucidating feature importance, contributing to critical clinical insights in the domain of healthcare monitoring for COVID-19.
Nordin et al. \citep{nordin2023explainable} propose an explainable predictive model for suicide attempts, combining complex ensemble learning models, such as Random Forest and Gradient Boosting, with the \textit{SHAP} method. This approach aims to enhance reliability in predicting suicide attempts while offering insights into critical factors for clinical decision-making. In the domain of mental health, the study highlights the superiority of Gradient Boosting with SHAP over Random Forest in terms of accuracy, underscoring the value of explainable machine learning techniques in healthcare.

\hl{Ahmed et al.{~\citep{ahmed2024explainable}} introduce a hybrid adaptive ensemble that merges several classifiers and incorporates XAI components to produce transparent, feature-based explanations. Their approach employs post-hoc feature-importance methods and local explanation techniques, allowing analysts to identify which network-traffic attributes most strongly influenced a given prediction.
The study in{~\citep{e2024evaluating}} addresses a major barrier to clinical AI adoption: determining whether current visual XAI tools genuinely support radiologists in reliably understanding model decisions. Instead of proposing a new method, the authors conduct a human-centered empirical assessment of widely used techniques, such as Grad-CAM, saliency maps, and other attribution visualizations, applied to chest X-ray classification. Their findings show that XAI visualizations vary in interpretability across different pathologies.
Tahir et al.{~\citep{tahir2024novel}} present a hybrid XAI framework that combines LIME and SHAP to deliver instance-level interpretability while preserving the global coherence offered by Shapley-based explanations. This work represents a meaningful advance toward practical interpretable AV systems through the use of hybrid XAI strategies.
Vani et al.{~\citep{vani2025personalized}} propose a deep learning approach for personalized health monitoring, employing SHAP to provide patient-specific interpretability. This enhances model transparency and promotes clinically grounded trust.
Alamro et al.{~\citep{alamro2025enhanced}} develop an intrusion-detection system based on dimensionality reduction and attention-driven deep learning, integrating XAI to improve threat detection and support decision-making. SHAP is used to deliver reliable interpretive insights within AI-enabled security environments.
Ghosh{~\citep{ghosh2025novel}} combines XAI with a deep learning model to examine large-scale financial datasets. By capturing temporal patterns in network activity, logs, and audit trails, the system identifies fraudulent financial transactions. LIME is employed to ensure transparency and evaluate the model’s interpretability.}

\subsubsection{Summary}

\hl{Recent research shows that XAI is essential for improving transparency, trust, and reliability in fields such as healthcare, security, finance, the IoT, NLP, and multimodal deep learning. Many studies focus on developing explanation frameworks, like LIME, Anchors, and SHAP, to make complex models easier to understand and more useful. A key area of research adapts XAI to specific challenges. In healthcare, for instance, visual and feature-based explanations support medical diagnoses and COVID-19 screenings. In security, XAI helps detect threats and analyze malware. Tailored methods also apply to self-driving cars and stock prediction, demonstrating XAI's ability to validate models and assist experts. Another important insight is that making models understandable can improve them. By visualizing key features or providing simple rules, studies often identify issues like model weaknesses and biases. This leads to stronger and more ethical models. User assessments show that the effectiveness of explanations relies on task complexity, model design, and user expertise. High-quality explanations need to be easy to understand, trustworthy, and practical. Overall, this review emphasizes progress in creating clear, reliable, and specialized AI systems, with hybrid methods and specific adaptations offering promising paths for future research.}


Distribution of the papers, reviewed in this section, by source and year are shown in Figure \ref{fig:sources_years_XAI} and distribution into the techniques used and their types is shown in Table \ref{table:techniques-10-XAI}.

\begin{figure*}[!ht]
	\centering
	\begin{subfigure}{.48\textwidth}
		\begin{tikzpicture}[scale=0.85]
			\pie[rotate=0,color={color1, color2, color3, color4, color5, color6, color7, color8, color9, color10, color11, color12}]{
				13/Elsevier, 
                5/PLOS, 
				17/IEEE, 
                4/Springer, 
				5/ACM, 
				8/ArXiv, 
				5/AAAI, 
				17/Nature, 
				5/PeerJ, 
				13/MDPI, 
				4/Hindawi, 
				4/ISMIR 
			} 
		\end{tikzpicture}
	\end{subfigure}
	\begin{subfigure}{.48\textwidth}
		\centering
		\begin{tikzpicture}[scale=0.85]
			\pie[rotate=0,color={color1, color2, color3, color4, color5, color6, color7, color8, color9, color10}]{
				13/2025, 
				13/2024, 
                4/2023, 
				13/2022, 
				9/2021, 
				22/2020, 
				4/2019, 
				9/2018, 
				9/2017, 
				4/2016 
			} 
		\end{tikzpicture}
	\end{subfigure}
	\caption{\hl{Distribution of the articles by source and year reviewed in section {\ref{sec:last-10-XAI}}.}}
	\label{fig:sources_years_XAI}
\end{figure*}

\begin{table*}[!ht]
	\caption{\hl{Distribution of the XAI works reviewed in section {\ref{sec:last-10-XAI}} into the techniques used and their types.}}
	\setlength{\tabcolsep}{5pt}
	\renewcommand{\arraystretch}{1.5}
\begin{center}
	\begin{tabular}{l | C{6.5cm} | C{1.5cm} | C{1.5cm} | C{1.5cm}} \hline\hline
		\multirow{2}{*}{Technique} & \multirow{2}{*}{Articles} & \multicolumn{3}{c}{Types} \\ \cline{3-5}
		&          & I / P & G / L & S / A \\ \hline
		SHAP & \citep{gramegna2020buy}, \citep{parsa2020toward}, \citep{debjit2022improved}, \citep{nordin2023explainable}, \citep{tahir2024novel}, \citep{ahmed2024explainable}, \citep{vani2025personalized}, \citep{alamro2025enhanced} & P & G & A \\ \hline
		LIME & \citep{mishra2017local}, \citep{feichtner2020understanding}, \citep{szczepanski2021new}, \citep{gite2021explainable}, \citep{tahir2024novel}, \citep{ahmed2024explainable}, \citep{ghosh2025novel} & P & L & A \\ \hline
		Grad-CAM & \citep{joo2019visualization}, \citep{sutton2022artificial}, \citep{naeem2022explainable}, \citep{e2024evaluating} & P & L & A \\ \hline
		Saliency map & \citep{fong2017interpretable}, \citep{e2024evaluating} & P & L & A \\ \hline
		Anchors & \citep{ribeiro2018anchors}, \citep{szczepanski2021new} & P & L & A \\ \hline
		Grammar evolution & \citep{cavaliere2020parkinson} & P & G & A \\ \hline
		LORE & \citep{guidotti2018local} & P & L & A \\ \hline
		Decision Trees & \citep{hara2016making} & P & G & S, A \\ \hline
		Cluster-TREPAN & \citep{de2020explainable} & P & L & A \\ \hline
   \end{tabular}
\end{center}
	\begin{tablenotes}
		\centering
		\item I = Intrinsic, P = Post-hoc, G = Global, L = Local, S = Model-specific, A = Model-agnostic
	\end{tablenotes}
	\label{table:techniques-10-XAI}
\end{table*}

\subsection{Cyber Forensics (CF)}\label{sec:CF}
A formal definition of CF is given in section \ref{sec:defs}. This section explores in more detail of why, what, and how of CF. Every contact by a perpetrator leaves behind traces \citep{chisum2000evidence}. To make a case against the perpetrator, these traces or pieces of evidence need to be found, collected, secured, studied, and analyzed. \textit{Cyber Forensics} \citep{alam2022cybersecurity,sammons2012basics,garfinkel2010digital} uses scientific methods and expertise to gather and analyze pieces of evidence found in cyber devices that can be used in criminal or other investigations in a court of law. This evidence can be used for different purposes, such as \textit{electronic discovery}, \textit{intelligence}, and \textit{administrative}. For example, the data collected from digital devices can provide actionable intelligence. This intelligence can help accomplish different types of missions, such as securing national interests, decreasing or eliminating crimes like kidnapping and child exploitation, etc. Electronic discovery is the process of searching, finding, and securing any electronic data to be used in a civil or criminal forensic case.

The main purpose of a CF process is to better understand and draw a conclusion about an event of interest by finding and analyzing the details and facts related to that event. In general this process comprises of four phases \citep{kent2006guide} as shown in Figure \ref{fig:CF_phases} and explained below:

\begin{figure*}[!ht]
	\centering
	\includegraphics[scale=0.53]{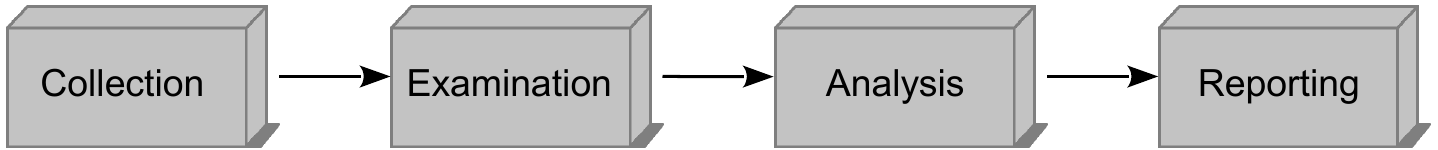}
	\caption{Four phases of Cyber Forensics.}
	\label{fig:CF_phases}
\end{figure*}

\begin{enumerate}
	\item
	\textit{Collection} -- Evidence collection is the foremost piece of work during CF. While collecting the data, there is a strong need of also preserve and secure that data. The evidence collected must be protected from accidental or intentional compromise. Cyber devices must be made inaccessible as soon as there is a guarantee that no volatile data will be lost. Collection is also performed promptly because of the dynamic nature of the data. The data may get lost in battery-powered devices and network connections if not collected promptly. Actions during the collection phase must be well documented. A forensic clone or image is made of the suspected cyber device, and the exam is conducted on the cloned device. The chain of custody must be maintained and documented. A proper Chain of Custody is to be maintained to ensure that the evidence collected has not been altered or changed during the forensic process.
	\item
	\textit{Examination} -- This is the process of forensically scrutinizing the data (small or large) using automated and manual methods/techniques. During this process, data of particular interest is extracted and analyzed, meanwhile preserving the integrity of the data. Moreover, the data files are defined that contain the information, which may be compressed, encrypted, and/or protected by access controls, which makes a meaningful extraction difficult. With the availability of AI techniques and tools, sifting through and extracting useful data has become much easier. It also reduces the amount of data that needs to be analyzed. To trust such AI processes/techniques, they need to be made explainable, and this makes the application of XAI in CF more compelling.
	\item
	\textit{Analysis} -- In this phase results of the examination are analyzed, using legally justifiable methods and techniques. The purpose is to extract useful information from the data that can be used and presented in a court of law. Different techniques are used, such as correlation of data, abnormality in data, such as slow network traffic, etc., and sifting through the data using automated (AI, ML, data mining, etc.) and manual techniques to predict an attack. Here, the use of AI becomes extremely beneficial. Not only does it automate (for now, partially) this process, but it also helps in extracting meaningful information from the data. But to accept such a piece of information in a court of law or the public, there is a strong need to make the process of extracting such information explainable, and this makes the application of XAI in CF more indispensable.
	\item
	\textit{Reporting} -- This is the last phase but also the most important phase of a forensics process. It makes the results of the analysis phase more comprehensive and understandable by both a technical and a non-technical person. It generally consists of the actions used during the forensic process, how and what tools/methods were used, and what other actions need to be performed, and may also provide recommendations and guidelines to improve the forensic process. It is challenging to provide an exact explanation, but one should provide as much detail as possible to the readers to help them make a responsible and accountable conclusion and decision about an event. Here, XAI can play a major role in providing human-understandable explanations that can be used by the forensic analyst to produce such a report.
\end{enumerate}

CF is an important tool for solving crimes committed against people where the evidence resides on a cyber device \citep{garfinkel2010digital}. Initially, forensic techniques were developed primarily for recovering data. As the techniques developed, network and memory forensics became possible, and CF grew widespread and reliable. Now, with the rise of complex cyber devices such as smartphones, IoT, etc., the proliferation of operating systems and file formats, pervasive encryption, use of the cloud for remote processing and storage, and legal challenges, CF is facing many new challenges. For example, there are dozens of systems running on smartphones, each with more than a million downloadable applications. File carving is one of the techniques to extract and classify raw data from a cyber device, e.g., a smartphone, into files. Sifting through such a large amount of raw data for classification to make sense of this data requires new techniques, such as from the field of AI \citep{alam2023sift}. To apply these techniques successfully in CF we need to justify and explain the results to a human \citep{hall2022explainable}. Here, the question arises: Does XAI provide such explainable results? This is the main focus of the study carried out in this paper.

\subsubsection{Classifications}\label{sec:CF_classifications}

The basic CF techniques are the same, but others are only applied in a certain environment, such as mobile, IoT, etc. Based on these working environments during the forensic process, in this section, we discuss some of the main classifications of CF. Every cyber device is kind of a computer device, i.e., it is capable of processing and storing electronic data. Therefore, we do not consider computer forensics as a separate class in CF. In this paper, we name this class as \textit{General} and we do not explain this class exclusively here.

\begin{enumerate}
	\item
	\textit{Mobile Forensics} -- Nowadays, we cannot imagine our life without a smartphone. These devices are mini-computers capable of almost the same functionality as a desktop computer. The number of mobile devices is growing exponentially. Their hardware and operating systems (OSs) are also evolving rapidly. Evidence can be located not just on the device or memory but also on the cellular network. Any component (base station, mobile switching center, home, visitor location register, etc.) in the network can potentially provide information that can be used in a forensic investigation. OS is responsible for creating and storing evidence and has a significant impact on device forensics. There are several mobile OSs, but the popular ones are Android, iOS, and Windows CE. Some of the potential items for collecting evidence in mobile devices are pictures and videos, deleted text messages, browser history, location information GPS, contacts, chat sessions, etc. The proliferation of mobile devices, each with a different OS, makes the forensics of these devices challenging. Also, their migrating nature makes it difficult to find the location of the device.
	\item
	\textit{IoT Forensics} -- IoT devices are all over the Internet, such as in smart cities, healthcare, smart offices/homes, vehicle automation, etc. The IoT market will continue to experience exponential growth. Like mobile forensics, IoT forensics is also facing a lot of challenges \citep{stoyanova2020survey}. A comprehensive list \citep{nist2014nist} is provided by the National Institute of Standards and Technology (NIST). We discuss here some of the important challenges. Due to the highly dynamic nature of the communication, it is almost impossible to know the number of devices to be confiscated and the scope of the damage. It is quite possible that important data, such as registry entries or files, could be completely erased before it is collected. Like mobile devices, due to their moving/migrating nature, locating the evidence becomes a challenge. The proliferation of these devices makes the forensics of these devices difficult. In IoT, the hardware and software used are heterogeneous. Maintaining a proper chain of custody in IoT is difficult. The reason is the presence of evidence data on multiple remote servers as in the cloud. Some IoT devices have a very limited memory and can easily be overwritten, resulting in the possibility of missing the evidence. One of the advantages of IoT forensics is that the evidence, if mirrored to multiple locations, as in the cloud, is harder to destroy by the criminals.
	\item
	\textit{Network Forensics} -- Application of scientific methods to investigate network breaches and vulnerabilities. These methods rely on discovering, capturing, identifying, and analyzing network traffic. Current network forensic mechanisms involve logging, packet marking, and heuristics based on various events of the network. Some of the major challenges of network forensics \citep{khan2016network,qureshi2021analysis} are. Due to the high network speed, it is difficult to capture packets and also requires a large amount of space to store packets for later analysis. Network intruders use several techniques, such as spoofing, to hide their IP addresses. As such, identifying the original IP address of the attacker becomes very difficult for forensic investigators. Like IoT forensics, due to the virtualized and distributed nature of networks, it becomes difficult to identify the location of the network devices to extract data. Due to the frequent mobility of data, the trust and integrity become low. That means the network may not have consistent, complete, and accurate data, which becomes a major concern for forensic investigators. Cloud is a wide area network and is used to host users and resources that can communicate using cloud-based technologies. Cloud forensics is part of network forensics and faces similar challenges and we do not consider this as a separate class in CF.
	\item
	\textit{Database Forensics} -- Any practical/working digital application contains a database of some kind. As an application becomes more complicated, a database is used to store important and sensitive information. Database forensics deals with the detailed analysis of a database, including its log files, metadata, and data files \citep{chopade2019ten}. Database forensics is an upcoming class of CF, and not much work has been done in this area. This is due to the inherent complexity and multidimensional nature of databases. Some of the major challenges of database forensics \citep{al2014towards} are. One of the basic requirements to perform successful database forensics is the understanding of the internal architecture of the database. A variety of database infrastructures (Oracle, MySQL, SQLite, DB2, etc.) and platforms (Windows, Android, UNIX, Solaris, etc.) make this task very difficult. Multidimensional databases can be viewed from different perspectives/features. This simplifies data presentation, navigation, and maintenance, but also makes it difficult to forensically analyze the tampering of such data \citep{fasan2012reconstruction}. As in other classes, scanning and making sense of large data composed of data files, logs, and metadata pose a challenge to performing successful forensics. Moreover, determining the location of tampering and suitable data in this large set of data, which may contain manual vulnerabilities and weaknesses created by inexperienced database administrators/users, poses resource and time constraints.
\end{enumerate}

\subsection{Last Ten Years of CF}\label{sec:last-10-CF}

To make readers familiar with the study carried out in the paper, we present a critical and short review of the last ten years of work carried out in CF. We exclude papers that explore the application of XAI in CF because we discuss in detail such applications in section \ref{sec:litreview}. A detailed survey on this topic is out of the scope of this paper. For a detailed survey, interested readers are referred to \citep{casino2022research}.

Nalawade et al. \citep{nalawade2016forensic} explore web browser forensics, focusing on collecting evidence from browsers like cache, history, cookies, and download lists. The study compares major tools for web browser analysis, emphasizing their benefits and limitations.
Anglano et al. \citep{anglano2016forensic} conduct a forensic analysis of ChatSecure, an Android Instant Messaging app, revealing vulnerabilities in its encrypted databases. Employing CF techniques, the study extracts the encryption passphrase from volatile memory, allowing decryption without user disclosure.
Zhang et al. \citep{zhang2017blockchain} present an approach, termed process provenance, to bolster the trustworthiness of the chain of custody in cloud forensics. By integrating blockchain and cryptography group signatures, the proposed method enhances proof of existence and privacy preservation for process records, addressing challenges arising from the distributed nature of cloud computing.
Koroniotis et al. \citep{koroniotis2019towards} introduce the Bot-IoT dataset, encompassing both normal and attack traffic, aiming to enhance the credibility of network intrusion detection and forensic systems. Statistical and machine learning methods are employed to evaluate the dataset's reliability, showcasing its potential for botnet identification in IoT-specific networks and providing a foundation for future network forensic models.
Nyaletey et al. \citep{nyaletey2019blockipfs} introduce BlockIPFS, aiming to address traceability and security concerns in the Interplanetary File System (IPFS) by integrating blockchain technology. BlockIPFS combines the efficiency of IPFS for file sharing with the traceability and security features of blockchain, offering a comprehensive solution for improved trustworthiness, authorship protection, and auditability in distributed file systems.

Dalezios et al. \citep{dalezios2020digital} focus on enhancing forensic capabilities in cloud platforms, addressing challenges tied to criminal activities. The Cloud Auditing Data Federation standard event model is employed to represent cloud-related events, promoting a clear and robust logging format for forensic investigations in cloud computing.
Rahman et al. \citep{rahman2020new} propose a web forensic framework that addresses the growing concern of automated bot crimes, providing a systematic approach. By leveraging real institutional web application logs and experimental bot attack scenarios, the authors develop unique access loggers in JavaScript and PHP to extract crucial patterns for forensic analysis.
Noura et al. \citep{noura2020distlog} propose a technique to countermeasure the anti-forensics methods that target the logs. The proposed technique first aggregates, compresses, and encrypts the logs. These logs are then fragmented, authenticated, and distributed over several storage nodes and secure the log files.
Luo et al. \citep{luo2020automatic} propose a framework for automatically inspecting and identifying Android applications that are inappropriate for kids under 12 years of age. Forensic information/content is extracted by parsing the HTTP packets. The contents are then classified as appropriate or inappropriate and a normal or abnormal report is generated, respectively.
Hina et al. \citep{hina2021sefaced} present SeFACED, an approach for email content classification. Outperforming traditional machine learning and deep learning models, SeFACED achieves robust accuracy and precision, contributing significantly to forensic investigations and enhancing the identification of malicious activities in email content analysis.

Shahbazi et al. \citep{shahbazi2022nlp} integrate natural language processing (NLP) techniques with a blockchain framework to enhance the security and performance of online social media investigations. By applying the Random Forest algorithm and blockchain technology, the study addresses system security and traces process changes, offering the potential for advancements in investigating cybercriminal activities and fraud.
Kim et al. \citep{kim2022methods} present a method for recovering deleted data from a Realm database. This database is mostly used and optimized for mobile devices. The method is based on the database structure and type. Such a method can also be used against anti-forensics techniques, such as deleting evidential data from the database, used by adversaries.
Gupta et al. \citep{gupta2023digital} explore the digital artifacts left by the Discord social media application on the Google Chrome web browser, focusing on Windows 10. The study employs various techniques, including the examination of local storage, log files, and the Google Chrome cache, offering insights for digital forensic analysts and researchers in the domain of social media forensics.
Nissan et al. \citep{nissan2023database} present a framework for predicting database query activity from a memory snapshot. The authors use a Support Vector Machine for such a prediction. This memory forensics allows forensic investigators to detect malicious activities even in the presence of incomplete logs.
Dragonas et al. \citep{dragonas2023iot} propose a forensic model to identify interconnectivity between IoT. IoT devices are connected to other IoT devices, and identifying this connectivity of things is essential to conducting a thorough investigation. A tool is also presented for extracting and visualizing this connectivity.
Domingues et al. \citep{domingues2023post} present a study for identifying and analyzing CF artifacts available in a post-mortem examination of a popular Zepp Life (monitors health and other activities) application for Android. The CF artifacts from these devices have been used in various criminal investigations \citep{domingues2023post}.

\hl{Mei et al.{~\citep{mei2024novel}} introduce a network-forensics framework that applies deep learning to attribute Advanced Persistent Threats (APTs)—that is, to infer an attacker’s identity, group affiliation, or campaign connections—from network telemetry. They aim to advance beyond simple detection and toward richer attribution and contextual forensic intelligence by learning patterns from large-scale traffic data, flow characteristics, and temporal behaviors.
Alqahtany et al.{~\citep{alqahtany2024integrating}} present an integrated blockchain and deep-learning model designed to enhance forensic traceability and preserve evidence integrity in Mobile VPN environments. Blockchain is employed to provide tamper-resistant logging, chain-of-custody verification, and distributed provenance, while the deep-learning component evaluates Mobile-VPN traffic and device artifacts to flag anomalies, identify misuse, and reconstruct session activity.
Sharma et al.{~\citep{sharma2025forensicllm}} propose ForensicLLM, a locally deployable large language model tailored for digital-forensics workflows. In contrast to cloud-based LLMs, ForensicLLM focuses on privacy, offline usability, reproducibility, and specialized reasoning for tasks such as log examination, artifact interpretation, and incident triage.
Kao et al.{~\citep{kao2025accelerating}} describe an NLP-driven system aimed at detecting cryptocurrency seed phrases (mnemonics) across multilingual datasets. Because such mnemonics appear in diverse linguistic forms, conventional pattern-matching techniques are insufficient; the authors therefore apply language-modeling methods to automatically identify, categorize, and extract mnemonic phrases within text sources encountered during cybercrime investigations.}

\subsubsection{Summary}

\hl{The reviewed literature highlights notable advancements in digital forensics, focusing on improved methods for gathering evidence, tracking it, and analyzing it automatically. Many studies concentrate on specific applications like web browsers, encrypted messaging apps, mobile databases, social media, health-monitoring apps, and Android apps. These advancements help investigators recover hidden, deleted, or encrypted data and understand user activity. Another important area is managing data and ensuring log integrity. Many papers use blockchain and tracking methods to secure logging systems and maintain evidence custody, especially in cloud services and social media investigations. Memory forensics also helps detect malicious actions when logs are incomplete. The integration of machine learning, natural language processing (NLP), and deep learning supports tasks such as intrusion detection and email classification. Specific datasets and frameworks, like Bot-IoT and ForensicLLM, enhance automation and consistent analysis. Lastly, research on IoT connections and bot activity detection emphasizes the need for modern investigations to handle complex environments. Overall, these studies demonstrate progress toward more reliable, automated, and integrity-focused digital forensics, addressing new challenges in cloud systems, mobile apps, IoT networks, and cybercrime.}


Distribution of the articles, reviewed in this section, by source and year are shown in Figure \ref{fig:sources_years_CF} and distribution into the forensics types (classes) is shown in Table \ref{table:types-10-CF}.

\begin{figure*}[!ht]
	\centering
	\begin{subfigure}{.48\textwidth}
		\begin{tikzpicture}[scale=0.85]
			\pie[rotate=0,color={color1, color2, color3, color4}]{
				55/Elsevier, 
				35/IEEE, 
				10/MDPI 
			} 
		\end{tikzpicture}
	\end{subfigure}
	\begin{subfigure}{.48\textwidth}
		\centering
		\begin{tikzpicture}[scale=0.85]
			\pie[rotate=0,color={color1, color2, color3, color4, color5, color6, color7, color8, color9, color10}]{
				10/2025, 
				10/2024, 
				20/2023, 
				10/2022, 
				5/2021, 
				20/2020, 
				10/2019, 
				5/2017, 
				10/2016 
			} 
		\end{tikzpicture}
	\end{subfigure}
	\caption{\hl{Distribution of the articles by source and year reviewed in section {\ref{sec:last-10-CF}}}}
	\label{fig:sources_years_CF}
\end{figure*}

\begin{table}[!ht]
	\caption{\hl{Distribution  of the CF works reviewed in section {\ref{sec:last-10-CF}} into the forensics types (classes) discussed in section {\ref{sec:CF_classifications}}}.}
	\setlength{\tabcolsep}{5pt}
	\renewcommand{\arraystretch}{1.5}
\begin{center}
	\begin{tabular}{l | C{5.5cm}} \hline\hline
		Types & Articles \\ \hline
		General & \citep{rahman2020new}, \citep{hina2021sefaced}, \citep{shahbazi2022nlp}, \citep{gupta2023digital}, \citep{sharma2025forensicllm}, \citep{kao2025accelerating} \\ \hline
		Network & \citep{nalawade2016forensic}, \citep{zhang2017blockchain}, \citep{nyaletey2019blockipfs}, \citep{dalezios2020digital}, \citep{mei2024novel} \\ \hline
		Mobile & \citep{anglano2016forensic}, \citep{luo2020automatic}, \citep{domingues2023post}, \citep{alqahtany2024integrating} \\ \hline
		IoT & \citep{koroniotis2019towards}, \citep{noura2020distlog}, \citep{dragonas2023iot} \\ \hline
		Database & \citep{kim2022methods}, \citep{nissan2023database} \\ \hline
	\end{tabular}
\end{center}
	\label{table:types-10-CF}
\end{table}

\section{Explainable Artificial Intelligence in Cyber Forensics (XAI-CF)}\label{sec:XAI-CF}

Here, we make a case for the significance and advantages of XAI in CF and strongly recommend its use in CF.

\subsection{Why XAI-CF}\label{sec:why-XAI-CF}
The ability of an AI system to learn and adapt without requiring user input makes it beneficial in automating and completing different tasks in several fields of work. Also, its ability to deal with large amounts of data makes it ideal to be used in CF, which deals with massive volumes of data \citep{zawoad2015digital}. Automation of evidence processing using AI-based techniques shows promising results in expediting the CF analysis process \citep{du2020sok}. Because of the inherent complexity of AI systems, one of the main problems with them is that they are not easy to understand and interpret, and it becomes difficult for humans to realize how an AI system came to a decision. This problem becomes more consequential when these systems are used in making important decisions, such as deciding a criminal case in a court of law, where pieces of evidence are forensically analyzed and determined through an AI system. Here we raise an important question and issue of \textit{how to build AI models that will be interpretable and explainable not only to the experts but also to users without prior knowledge or expertise in the area of CF}.

In general, the stakeholders of CF are not experts in AI. For them to make an informed decision, they need to first understand the output of the respective AI system. The use of XAI in CF improves the CF analysis and extracts forensically important pieces of evidence in a way that is easy to interpret and understand by the different stakeholders of CF, and hence can be used successfully in investigations and potentially in a court of law. We are already stepping into the digital age, and if we want CF to take full advantage of this and successfully address the challenges \citep{karie2015taxonomy,du2020sok} it faces, then we need to fully infuse AI into CF processes. To successfully infuse AI into CF, we need to interpret and explain the results of an AI system to the stakeholders of CF. There are several studies 
\citep{hepler2007object,keppens2012argument,fenton2013general,timmer2014extracting,verheij2014catch,vlek2014building,timmer2015structure,verheij2016arguments,vlek2016method,aditya2018enabling,afzaliseresht2019explainable,costantini2020assessing,bolle2020role,mahajan2021explainable,pethe2022atle2fc,hall2022proof,jayakumar2022visually,de2022interpretable,bouter2023protoexplorer} 
carried out to explain AI outputs for the stakeholders of CF involved in the decision-making process. We discuss some of them in detail in section \ref{sec:litreview}. 

\subsection{How XAI-CF}\label{sec:how-XAI-CF}
When we talk about AI \citep{AI-modern-approach-2020}, two things come to mind: \textit{intelligence}, such as the machines that perform functions that require intelligence when performed by people, and \textit{automation}, such as decision-making, problem-solving, learning, etc. AI is a universal field \citep{AI-modern-approach-2020} and encompasses a large variety of subfields, methods, and techniques, such as knowledge representation, adversarial searching, statistical and probabilistic reasoning, supervised, unsupervised, and reinforcement learning, and natural language processing. In an AI system for CF, one or a combination of these techniques is used to perform CF analysis, and a decision (output) is reached. In an XAI system for CF, one or a combination of these techniques is also used to interpret and explain the decision (output) of the AI system. An XAI system can be divided into four categories \citep{doran2017does}, opaque, interpretable, comprehensible, and truly explainable, as shown in Table \ref{tab:XAI-categories}.

\begin{table*}[!ht]
	\caption{Four categories of an XAI-CF system}
	\setlength{\tabcolsep}{5pt}
	\renewcommand{\arraystretch}{1.3}
\begin{center}
	\begin{tabular}{ l | L{12.9cm} } \hline\hline
		Category         & Description \\ \hline
		Opaque            & The mappings between input and output are invisible; a proprietary AI system where the licensor does not want to reveal the workings of the system; most difficult to interpret and explain. \\ \hline
		Interpretable     & The mappings between input and output are visible and understandable; a linear regression model where the feature weights can be compared to realize the relative importance of each feature; easier than opaque systems to interpret and explain. \\ \hline
		Comprehensible    & The system's output is tagged with symbols (words or visualization) that allow the user to relate properties of the inputs to their output; the degree of comprehension about the symbols depends on the user's knowledge about the AI system; a Bayesian network where edges are tagged with symbols to explain the transition; easier than interpretable systems to interpret and explain. \\ \hline
		Truly Explainable & This is both interpretable and comprehensive; moreover, it provides reasoning that explains the decision-making process using human-understandable features of the input data; a simple decision tree where edges are tagged with symbols to explain the transition, and also provides formal and informal reasoning for each decision taken in the process; most easy to interpret and explain, and are human-understandable. \\ \hline
	\end{tabular}
\end{center}
	\label{tab:XAI-categories}
\end{table*}

The first three categories encompass the current work on XAI-CF. The last category is very difficult to achieve in XAI-CF because of the critical nature of the CF processes and the inconsistent reasoning process of humans to understand a decision. A human-in-the-loop \citep{nguyen2021human,solanke2022explainable} may help an XAI-CF model achieve true explainability. Here, we discuss some of the techniques to achieve XAI-CF. Table \ref{tab:lit-review-summary} highlights these techniques in current and previous works on XAI-CF discussed in section \ref{sec:litreview}.

\subsubsection{Model Simplification}
One of the basic approaches used by XAI models is to simplify the original model. A simplified model is easy to interpret, understand, and explain. There are different techniques to achieve this. One of the techniques used is a surrogate model \citep{aditya2018enabling}. Using the predictions of the original model, a surrogate model is built, such as a simple decision tree. This transparent model is built when it is difficult to know the influence of the features on the prediction, and is trained to approximate the behavior of the original model. The surrogate model helps understand the decision-making process of the original model. One of the problems with these surrogate models is that they may provide the same predictions as the original model but with different rationales and reasoning. This could produce contrasting relations among pieces of evidence that can lead forensic analysts and other stakeholders of CF to reach a different decision than without a surrogate model. To address this limitation, researchers are working \citep{mariotti2023beyond} on developing metrics to evaluate faithful surrogate models in XAI. Another technique used is storytelling \citep{afzaliseresht2019explainable} that highlights the semantic and inferred information from the model in a human-readable format. Other techniques \citep{verheij2016arguments} used for formalizing and simplifying reasoning about the forensic legal evidence model are narrative, argumentative, and probabilistic approaches.

\subsubsection{Feature Importance/Relevance}
The most popular method used in XAI-CF is extracting the features, building the model, and then interpreting and explaining the model by computing the importance and relevance of the input features to the output of the model. Having this knowledge helps in crafting different inputs to the model for testing the robustness of the model and hence increases trust in the model. Feature importance helps in explaining the model using different techniques that reduce the efforts of humans to interpret the model. Feature relevance also helps in selecting the best model for forensics analysis. One such method \citep{aditya2018enabling} first simplifies the model using a surrogate model and then computes the influence of the features on the prediction. \citep{afzaliseresht2019explainable} mines temporal patterns from the model and uses storytelling to relate patterns to output and highlight the information in a human-readable format. \citep{mahajan2021explainable} uses LIME \citep{ribeiro2016should} to relate the relevance of input features to the output and conclude that LIME is an important XAI technique to select the best model for forensics analysis.

\subsubsection{Visualization}
This technique, in our opinion, is the best approach to explain a model and is easy to understand by humans. It is highly desirable in an XAI-CF system to have an interactive, data-driven visualization tool for forensics analysis. This increases the transparency of a model and greatly improves the ability of a forensics analyst and other XAI-CF stakeholders to understand and reason about the model. This ability helps them in the final decision process. A saliency or a heat map visually explains the contributions of the model. Sometimes a graph of the model is tagged with symbols that relate the input to its output. Visualization of the output can greatly help in interpreting and explaining a model to a forensics analyst and other stakeholders of CF. One such work \citep{keppens2012argument} extracts argument diagrams (tagged with symbols) visually representing a reasoning structure from the model. \citep{bouter2023protoexplorer} visualize and interpret deep fake video data for forensics analysis. A forensics analyst can zoom in on a part of the image and observe the contributions of the model to the prediction score. As we can see, there is a lot of room for improvement in this area, and there is a need to develop new innovative methods to visualize CF models for better explanation, especially for non-experts. A human-in-the-loop can complement such a visualization to better understand and explain the model.

\subsubsection{Human-In-The-Loop}
A human-in-the-loop can complement an XAI-CF system to produce valuable, practical, and structured results.  In our opinion, this is the most vital technique to achieve truly explainable systems. There are three approaches to including humans in AI: \textit{Active Learning} \citep{settles2009active} -- where AI has more control than a human; \textit{Interactive Learning} \citep{amershi2014power} -- where there is closer interaction between a human and AI; and \textit{Machine Teaching} \citep{ramos2020interactive} -- where a human has more control than AI. For XAI-CF, we recommend using the middle approach the \textit{Interactive Learning}. One of the main reasons for this recommendation is the nature of the data available in CF. Most of the data in CF is in binary form, such as hard disk raw data for file carving, malware programs, images, video, etc. This kind of data falls into the category of either semi-structured or unstructured data. To convert this raw data into a suitable internal representation for AI to detect or classify patterns, interactive learning is useful in giving additional structure to this raw data. Another main reason is the explainability that interactive learning can provide in an XAI-CF system. In addition to being part of the learning process in XAI-CF, humans can be at the end, interpreting what the model has learned. Adding forensics experts to the learning helps XAI-CF to improve its knowledge and, in turn, its explanation. Interactive learning also comes with some shortcomings \citep{mosqueira2023human}, such as the intermingling of AI and human aspects and the dependence on a human (forensics) expert. Currently, there are not enough works that study the role of human-in-the-loop in XAI \citep{abdul2018trends}. Recently, in the area of XAI-CF, \citep{nguyen2021human} proposed to use an expert forensics analyst as a human-in-the-loop to provide adjustments so that with each iteration, the system improves and the result produces a better explanation of the model.

\subsection{Evaluating XAI-CF}
There are metrics for evaluating an AI system, and there is work in progress on developing metrics for evaluating the performance of an XAI system. How to measure the accuracy of an XAI system? There are certain criteria, such as how good (precise and clear) the explanation is, how satisfied it is for the respective audience is, and how trustworthy (justifiable) the explanation is. A detailed taxonomy of the XAI system's evaluation methods is given in \citep{lopes2022xai}.

Most of the previous works propose metrics that only work with a specific XAI model \citep{arras2022clevr,keane2021if,lakkaraju2017interpretable,nguyen2020quantitative,rosenfeld2021better,vilone2020comparative} or use humans to study and give their opinions about the usefulness of an XAI model \citep{dieber2022novel,holzinger2020measuring}. Only three approaches \citep{hoffman2018metrics,rosenfeld2021better,sovrano2023objective} claim to be model agnostic and can be used with any XAI-CF model. Here we briefly discuss these three approaches.

Hoffman et al. \citep{hoffman2018metrics} listed three major criteria to evaluate XAI: how precise and clear the explanation is; how satisfying it is for the respective audience, and how justifiable the explanation is. The authors recommend using researchers and experts to manually use these criteria to evaluate XAI models with the help of checklists, certain scales, and psychometrics.

Rosenfeld et al. \citep{rosenfeld2021better} introduce four objective metrics to quantify XAI. Given the XAI goal, these metrics quantify the explanation itself and its appropriateness. The first metric compares the performance of the black-box model with a transparent model. XAI-CF demands a binary evaluation because of the involvement of legal issues. This metric can be used to evaluate the XAI-CF model. Any value greater than a threshold nullifies the explanation. The second metric measures the complexity of the application by the number of rules used in the explanation. The more simplifier an explanation is, the more preferable it is. The third metric focuses on the number of features used to construct the explanation. The fourth metric measures the stability of the explanation. For all these metrics, a threshold value is required, which is computed by conducting user studies of the specific XAI model. 

Recently, Sovrano et al. \citep{sovrano2023objective} propose the degree (how much) of explainability as one of the metrics to measure the accuracy of the explanation described in natural language. The authors use pre-trained \textit{deep language models} for general-purpose answer retrieval \citep{bowman2015large,reimers2019sentence,yang2019multilingual,karpukhin2020dense} to measure explainability. A graph is extracted by detecting, with a dependency parser, all the clauses within the explanation's text. These clauses are represented as special triplets: subject, body, and object. A natural language representation is obtained of these triplets as a possible answer using pre-trained \textit{deep language models}. The degree of explainability is then quantified by measuring its relevance to answering a pre-defined set of conventional questions.

\subsection{Desiderata of XAI-CF}
\begin{figure*}[!ht]
	\centering
	\includegraphics[scale=0.5]{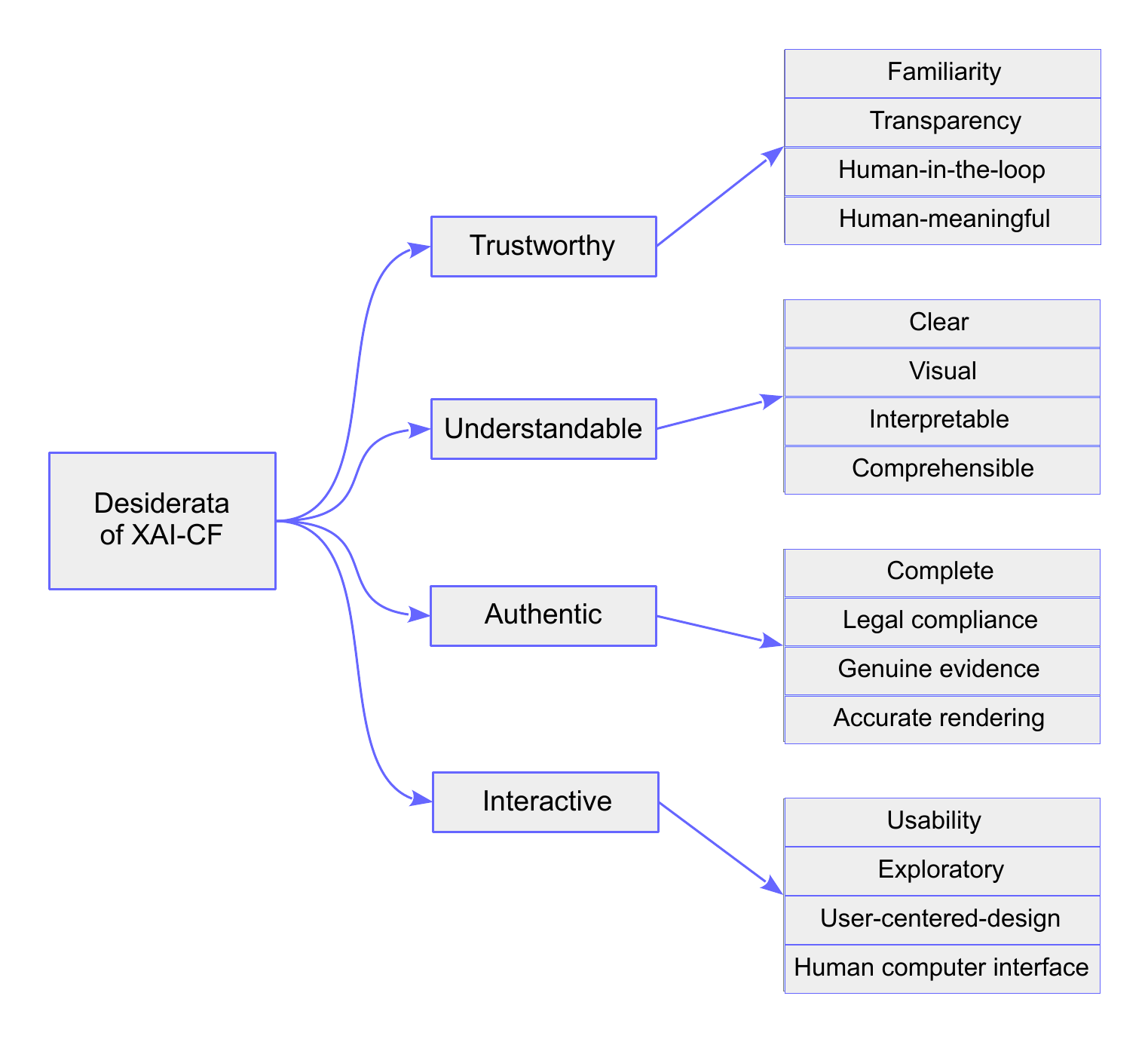}
	\caption{Desiderata of XAI-CF, divided into four main categories and sixteen sub-categories.}
	\label{fig:XAI_CF-desiderata}
\end{figure*}

Here we discuss some of the main requirements and prerequisites of a successful and practical XAI-CF system. There are many studies \citep{doran2017does,gunning2019darpa,phillips2021four,arrieta2020explainable,vilone2020explainable,lopes2022xai} that describe and discuss the prerequisites of a successful XAI system, such as \textit{Trustworthy}, \textit{Transparent}, \textit{Effective}, \textit{Efficient}, \textit{Interpretable}, \textit{Informative}, \textit{Clear}, \textit{Simple}, and so on. We take into consideration these prerequisites and add some of ours, and divide the requirements into four main categories and sixteen sub-categories as shown in Figure \ref{fig:XAI_CF-desiderata} that equip an XAI-CF model to become a successful and practical system.

\subsubsection{Trustworthy}
An XAI-CF system is used to make important and critical decisions, especially in a court of law. Before making such decisions based on the results and pieces of evidence provided by an XAI-CF system, stakeholders of CF must trust the XAI-CF system. Therefore, this is one of the most critical prerequisites of an XAI-CF system. Trust in an XAI-CF system influences a user's ability to believe in the output of the system without any uncertainty \citep{bhatt2020explainable}. The foremost property that can increase the trust of a user in an XAI-CF system is the quality of explanations provided by the system. Generally, users trust explanations created by humans rather than machines \citep{kunkel2019let}. An explanation by a machine can be improved if we integrate \textit{human-in-the-loop} in an XAI-CF system. In this way, both of them complement each other and can produce more trustworthy results. This integration is still in its infancy and is one of the challenges we discuss later in the paper. A \textit{human-meaningful} explanation by an XAI-CF system also increases trust in the system \citep{nourani2019effects}. Here, the human-meaningful explanation means understandable by humans. As there are different stakeholders of CF, the explanation should be tailored to each group of stakeholders. The trust in an XAI-CF system increases as the user becomes more familiar and experienced with the system over its time of use \citep{holliday2016user,yang2017evaluating}. For an XAI-CF system to become trustworthy, this \textit{familiarity} and experience should be positive. The trust of the user on an XAI-CF system can be measured using interview questions, both closed and open-ended, and also Likert-scale \citep{albaum1997likert} questionnaires can be used to measure the user's trust. \textit{Transparency} \citep{felzmann2019transparency} is another property of an XAI-CF system that increases trust. Making information open and transparent requires that the user be literate enough to make use of the information provided and can assess the risks of XAI-CF decision-making systems. Different stakeholders of CF have different abilities to make use of the information provided; therefore, transparency should be tailored to the user's ability. For example, an XAI-CF system should be transparent enough to a forensics analyst so that they can extract the right pieces of evidence to help in making an informed decision. Similarly, the forensics analyst should be familiar enough with the XAI-CF system to complete such a task.

\subsubsection{Understandable}
An XAI-CF system should be understandable to the stakeholders of CF. \textit{Interpretability} is the ability to explain a model in understandable terms to humans \citep{doshi2017towards}. Understandability and interpretability are complementary. \textit{Clarity} and \textit{Comprehensibility} are the two other properties that an explanation should have to be more understandable, and all three are also complementary to understandability. Understandability depends on the perception of the user. The explanations provided by an XAI-CF system help a user build a mental model of how the system works. In cognitive psychology, mental models \citep{johnson1983mental} give a person the ability to reason and make decisions. \textit{Visualization} is one of the best approaches to explaining a system and easiest to understand. One of the methods that can be used to assist humans in building a better mental model of an XAI-CF system is to represent the system, including explanations, graphically/visually, such as causal loop, stock, and flow diagrams \citep{groesser2012mental}. Causal loop diagrams (CLDs) \citep{haraldsson2004introduction} can be used to explicitly map an XAI-CF system and make it transparent and visible to others. A CLD consists of three basic elements: boxes, connections, and feedback loops. CLDs aid in analyzing a system qualitatively. To perform a quantitative analysis, a CLD is transformed to a stock and flow diagram \citep{sterman2001system}. All these visualization techniques are being used in a wide range of areas to understand the system dynamics and we recommend using such techniques to help increase the understandability of an XAI-CF system. There is not much work done in building better mental models to increase the understandability of an XAI system. This area is open to a lot of future work. For example, such an XAI-CF system can use boxes/nodes to represent input features and output results, connections/edges can represent the correlation or influence of input features to output results, and feedback loops can represent different iterations in the system. We can use tags/labels both at nodes and edges to improve the flow of information for better understanding. In the future, we intend to work in this area and explore this idea further by designing and developing such an XAI-CF system.

\subsubsection{Authentic}
An explanation is authentic when it \textit{accurately renders} and expresses a model's behavior \citep{markus2021role}. To understand a model we need to know how the model behaves under different conditions. The explanation of this behavior needs to be rendered correctly without any bias. Another property that makes an XAI-CF system authentic is \textit{completeness}. There is the famous quote \say{Justice is blind}, but as we know, sources of information can bias the judicial process \citep{ahola2010justice,franklin2018state}. The pieces of evidence produced by an XAI-CF system should be unbiased and complete. Bias can occur in an AI system for different reasons: a human-in-the-loop in an AI system; the inability to generate training data that is truly representative of the entire population of an AI system, etc. It is one of the major challenges of AI to keep the system unbiased. It may not be possible to remove all the biases; therefore, whenever a piece of evidence is presented that is generated by an XAI-CF, we should explain the biases in the system for the stakeholders to make an informed decision. Authenticity also requires an XAI-CF system to adhere to the \textit{legal compliance} guidelines to make the information admissible in legal proceedings. For example, we should pay attention to the legal rules surrounding the collection and use of digital evidence; otherwise, it makes the evidence worthless and can leave the investigators vulnerable to countersuits. A piece of forensic evidence must satisfy two main conditions: it must be relevant \citep{mason2001scientific}; and derived by scientific methods supported by appropriate validation \citep{giannelli2005daubert}. For more details and coverage of the legal aspects of CF, interested readers are referred to \citep{ryan2002legal}. One of the most critical rules of CF is not to change the original data. While collecting the evidence, forensic analysts use write blockers to follow this rule. Write blocking preserves the original evidence by preventing any data from being written to the original evidence device \citep{alam2022cybersecurity}. This ensures the collection of a \textit{genuine evidence}. For an XAI-CF system to be authentic, it needs to preserve the genuineness of the pieces of evidence.

\subsubsection{Interactive}
After an initial explanation is presented, a user may have follow-up questions/queries \citep{adadi2018peeking}. To let the user perform such an action, we need to allow the user to interact with the XAI-CF system. For example, in a forensic deep fake analysis of a video \citep{bouter2023protoexplorer}, a forensics analyst zooms in on a part of the video image to observe the contributions of the XAI-CF system to the prediction score, to decide if the video is manipulated or not. Arya et al. \citep{arya2019one} distinguish between static and interactive explanations. The first one does not change in response to feedback from the users, whereas the latter allows users to interact and drill down or ask questions until they are satisfied. XUI (Explanation User Interface) is a new concept and is defined as \say{the sum of outputs of an XAI system that the user can directly interact with} \citep{chromik2021human}. There are two modes of XUI \citep{shneiderman2020bridging}, explanatory, where it gives a single explanation (i.e., visual or text), \textit{exploratory}, where it lets users freely explore the AI model behavior. In this paper, we recommend the interactive explanation provided by the exploratory XUI as defined above. For an XAI-CF system to be interactive, it needs an easy-to-use \textit{Human Computer Interface} (\textit{HCI}). One of the basic goals of HCI is to improve the visual design for the interaction of users. \textit{Usability} \citep{punchoojit2017usability} can be applied to improve the visual design of an XAI-CF system. This is one of the major factors that affect the satisfaction of a user. Another important requirement for a good interactive system is \textit{user-centered-design} \citep{vredenburg2002survey}. This means that the system is designed based on an explicit understanding of the users, tasks, and environment. This is an iterative design process where in each iteration, the focus is on the users and their needs. A grand challenge of HCI is the design of interfaces that \say{allow users to understand the underlying computational processes better} \citep{shneiderman2016grand}. There is limited research reported on HCI and XAI \citep{wolf2019explainability,shneiderman2020bridging} and very few \citep{bouter2023protoexplorer} on HCI and XAI-CF. To build XAI-CF systems and make them successful in practice, there is a strong need to further research in this area and provide the user with a usable interactive XAI-CF system.

\section{Literature Review}\label{sec:litreview}

\begin{figure*}[!ht]
	\centering
	\includegraphics[scale=0.55]{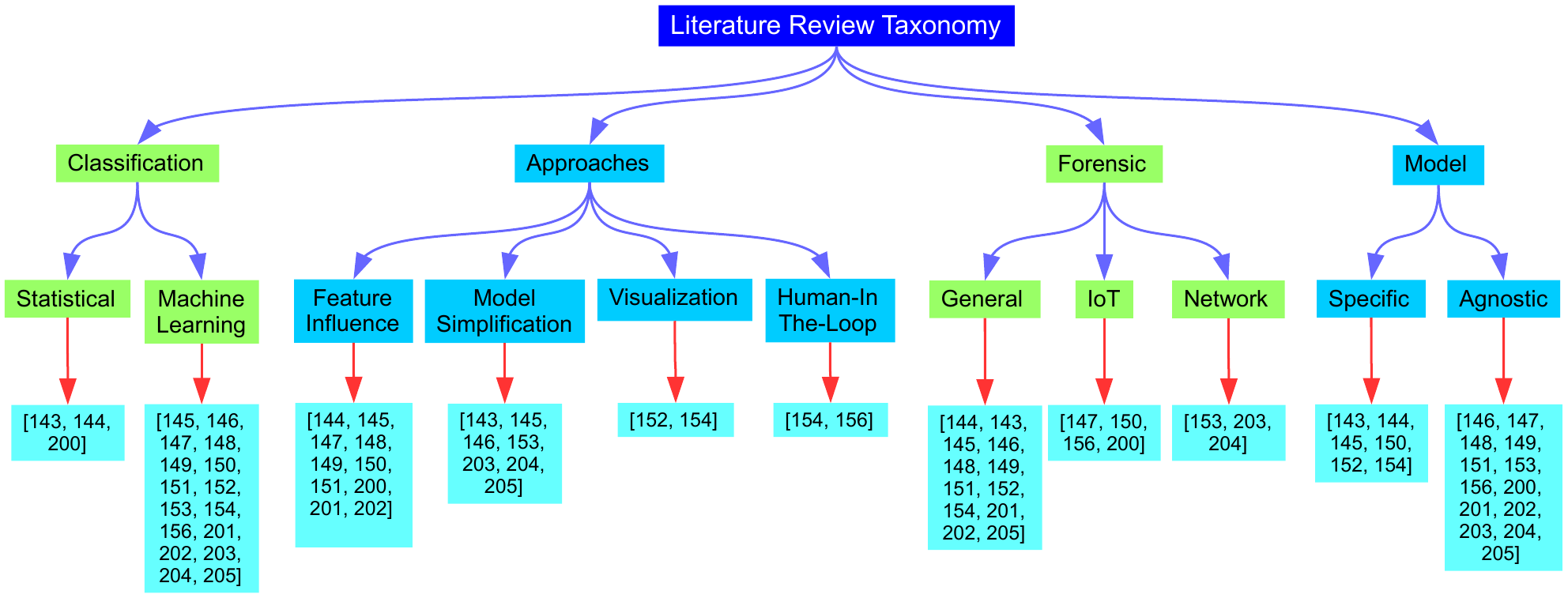}
	\caption{\hl{The taxonomy of the works discussed in section {\ref{sec:litreview}}. Some of the works overlap, because they cover more than one taxonomy.}}
	\label{fig:taxonomy_literature_review}
\end{figure*}

\hl{This section discusses and reviews the current works on the application of XAI in CF. We divide the work into two classes: \textit{Statistical} -- works that use statistical techniques to interpret or explain results of AI CF methods, and \textit{Machine Learning} (ML) -- Works that use ML techniques to interpret or explain results of AI CF methods. Figure{~\ref{fig:taxonomy_literature_review}} presents a complete taxonomy of the works that are reviewed in this section.}

\hl{Three articles are classified under statistical and were published between the years 2016 -- 2023, whereas twelve articles are classified under ML and were published between the years 2018 -- 2025. Most of the works employed the approach \textit{feature influence}, which shows its popularity and ease of implementation. The approach \textit{Model simplification} is also employed almost in the same proportion as the \textit{feature influence}, and it seems like it is also popular but difficult to implement. \textit{visualization} and \textit{Human-in-the-loop} approaches are the least employed because they are very new phenomena and difficult to implement, but hopefully will be more popular and employed in the future.}

		
		
		
		
		

\hl{A summary of the articles reviewed in this section is shown in Table{~\ref{tab:lit-review-summary}}. The research works reviewed are published in the last ten years (2016 -- 2025). Although the word XAI was coined in 2018{~\citep{gunning2019darpa}}, there are research works carried out before 2018 on explaining the statistical (part of AI) pieces of evidence presented in a legal case, which is part of the CF as per definition{~\ref{def:CF}} in this paper. Only two works use the state-of-the-art XAI method LIME to evaluate and explain AI models. Bayesian networks are popular for visualizing legal pieces of evidence and reasoning. Most of the statistical techniques are used to explain these Bayesian networks for the legal stakeholders. Out of these nineteen research works reviewed here, the most modern and advanced approaches employed are presented in{~\citep{bouter2023protoexplorer}} and{~\citep{khalid2024towards}} in the years 2023 and 2025. {\citep{bouter2023protoexplorer}} used visualization and a human-in-the-loop to interact and improve the XAI model with each iteration. {\citep{khalid2024towards}} presents a holistic framework to integrate XAI across the CF pipeline. This shows that the current recent work in this area is moving in the right direction. This effort is appreciated but relatively rare compared to other XAI applications in other domains. Therefore, we must promote and facilitate this process to keep pace with the ever-changing technology in the present digital age.}

\begin{table*}[!ht]
	\caption{\hl{Summary, forensic (as discussed in section {\ref{sec:CF_classifications}}) and model types, of XAI-CF research works reviewed in section {\ref{sec:litreview}}}.}
	\setlength{\tabcolsep}{3pt}
	\renewcommand{\arraystretch}{1.2}
\begin{center}
	\begin{tabular}{ c | c | c | c | L{13cm} } \hline\hline
		\multirow{2}{*}{Article} & \multirow{2}{*}{Year} & \multicolumn{2}{c|}{Types} & \multirow{2}{*}{Summary} \\ \cline{3-4}
		&                       & Forensic & Model  &  \\ \hline
		
		
		
		
		
		\citep{vlek2016method} & 2016 & G & S & Extend \citep{vlek2014building} by providing a probabilistic interpretation of scenario quality for explaining and understanding Bayesian networks with scenarios \\ \hline
		
		\citep{verheij2016arguments} & 2016 & G & S & Present a hybrid model that connects arguments and scenarios to improve coherent scenario-based interpretations that can assist in explanation \\ \hline
		
		\citep{khan2018eliciting} & 2018 & G & A & Associative rule mining to analyze security event logs and an action-based model is proposed for automatic examination for investigation purposes \\ \hline
		
		\citep{aditya2018enabling} & 2018 & G & S & Adversary testing framework to develop trust in a black box deep neural networks models in the context of CF \\ \hline
		
		\citep{afzaliseresht2019explainable} & 2019 & G & A & A contextualized interpretation of security events that can help explain the events and hence reduce the efforts of humans to interpret events\\ \hline
		
		\citep{costantini2020assessing} & 2020 & I & A & A framework that provides transparency to the reasoning process of evidence collection during forensic analysis of IoT devices for investigation purposes. \\ \hline
		
		\citep{bolle2020role} & 2020 & G & A & A framework based on scientific interpretation principles and formalizing different steps such as forensics evaluation to help evaluate, understand, and explain the outputs of automated AI CF system \\ \hline
		
		\citep{mahajan2021explainable} & 2021 & G & A & Apply LIME, one of the XAI techniques, to evaluate AI models for forensics analysis of toxic comments \\ \hline
		
		\citep{nguyen2021human} & 2021 & I & A & Integrates human (forensics expert) and machine in an interactive and iterative cycle that uses human intelligence to drive the XAI-CF system \\ \hline
		
		\citep{pethe2022atle2fc} & 2022 & I & S & A model with an explainable layer that consists of Frequent Pattern Growth, one of the XAI techniques, and a deep learning classifier \\ \hline
		
		\citep{hall2022proof} & 2022 & G & A & Apply LIME, one of the XAI techniques,  to evaluate AI models for explaining forensics analysis of image and video file content \\ \hline
		
		\citep{jayakumar2022visually} & 2022 & G & S & Apply Anchors, one of the XAI techniques, to provide visual explanations for deep fake detection for improving the interpretability of the model \\ \hline
		
		\citep{de2022interpretable} & 2022 & N & A & A federated transformer log learning model that provides forensic investigators with actionable information to effectively analyze system activity and operation \\ \hline
		
		\citep{gopinath2023explainable} & 2023 & I & A & Present metrics, such as the steps and time consumed in image creation and data retrieval, that can increase trust in the pieces of evidence in IoT devices \\ \hline
		
		\citep{bouter2023protoexplorer} & 2023 & G & S & The forensics analyst can explore and refine the visualized model and provide an explanation of how the model aided in making the decision \\ \hline

        \citep{alam2024sift} & 2024 & G & A & Optimizing file-type identification with integration of XAI methods, LIME and SHAP, to show which regions or features led to the predicted file type. \\ \hline

        \citep{khalid2024towards} & 2024 & N & A & A holistic framework that embeds XAI techniques, such as SHAP, LIME, and Anchors, etc., throughout a CF workflow. \\ \hline

        \citep{hermosilla2025explainable} & 2025 & N & A & Compares SHAP and LIME explanations when applied to two classifiers (XGBoost and TabNet) for network intrusion detection. \\ \hline

        \citep{khalid2025bridging} & 2025 & G & A & Explores unsupervised learning paired with XAI techniques for forensic domains that produce large volumes of unlabeled data. \\ \hline

    \end{tabular}
\end{center}
	\begin{tablenotes}
		\centering
		\item G = General forensics, I = IoT forensics, N = Network forensics, S = model-specific, A = model-agnostic
	\end{tablenotes}
	\label{tab:lit-review-summary}
\end{table*}

\subsection{Statistical}

Statistics plays a vital role in AI methods \citep{friedrich2022there}. This section presents a review and discussion of some of the research works that explain AI systems in CF using statistical techniques. Although there are several research works, such as \citep{hepler2007object, keppens2012argument, fenton2013general, timmer2014extracting, verheij2014catch, vlek2014building, timmer2015structure} done before 2016, we do not include them in the literature review because of the constraints of the published date.

Vlek et al. \citep{vlek2016method} proposed a method for explaining and understanding Bayesian networks with scenarios. The authors adapted a previously proposed scheme \citep{vlek2014building} for building a Bayesian network with scenarios. Scenarios in text form can be extracted from a Bayesian network and presented in a report with evidential support and information about the quality of the scenario. This makes the Bayesian network understandable to a legal expert. The paper mentions scenario quality and evidential support, indicating a need for understanding the strength of the presented scenarios. The quality of scenarios can be contributed by being explained in terms of probabilistic interpretations, and insights can be provided into how well each scenario is supported by the available evidence.

Verheij et al. \citep{verheij2016arguments} present a study of connections between arguments, scenarios, and probabilities and how they affect reasoning with pieces of evidence. In this study, an overview of combining the three approaches in different ways, such as pairwise and all three is presented and reviewed. This study helps in understanding the connections between probabilistic and non-probabilistic pieces of evidence that are useful in criminal investigation and decision-making. The authors claim that this will reduce reasoning errors and miscommunication between legal and forensics experts. The goal of the study is to reduce miscommunication between legal and forensic experts. Transparent and clear explanations of probabilistic information can play a role in achieving this goal. Also, the authors mention a hybrid model that connects arguments and scenarios. The incorporation of probabilistic elements into this hybrid model and the bridging between adversarial argumentative approaches and globally coherent scenario-based interpretations can be assisted in explanation.

Gopinath et al. \citep{gopinath2023explainable} perform forensic analysis of IoT devices in a method that increases trust in the analysis. They perform investigations on digital evidence and present certain metrics to increase trust in the pieces of evidence. These metrics are reports of different scenarios in detail, such as steps to retrieve the data, how the report was generated, etc, and the time consumed in image creation and data retrieval. These metrics are obtained using statistical methods for a comprehensive understanding and application of quantitative methods in these scenarios. These metrics require little effort on the part of the forensic analyst but can greatly help in the decision process of whether to trust the pieces of evidence collected or not.

\subsubsection{Summary}

\hl{The three papers focus on enhancing the interpretability and trustworthiness of forensic analyses through structured explanations. Vlek et al.{~\citep{vlek2016method}} propose scenario-based explanations for Bayesian networks, providing probabilistic interpretations and evidential support to make the networks understandable to legal experts. Verheij et al.{~\citep{verheij2016arguments}} explore hybrid models that connect arguments, scenarios, and probabilities, aiming to reduce reasoning errors and miscommunication between legal and forensic experts. Gopinath et al.{~\citep{gopinath2023explainable}} focus on IoT forensics, presenting quantitative metrics and detailed scenario reports that increase trust in digital evidence while requiring minimal effort from analysts. Collectively, these works highlight the importance of transparent, structured, and quantitative explanations to support reliable forensic decision-making.}


\subsection{Machine Learning (ML)}

Machine Learning (ML) is a major and core part of an AI system. ML gives intelligence to machines and makes them partially think on their own. There are three basic approaches to ML, supervised, unsupervised, and reinforced learning. Here we are not going to go into the details of these approaches. For an introduction and a detailed discussion on ML, interested readers are referred to \citep{Introduction-ML-2020}. This section presents a review and discussion of some of the research works that explain AI systems in CF using ML.

Khan et al. \citep{khan2018eliciting} present an associative rule mining technique to analyze security event logs, which can be used during forensics investigation of security events. Firstly, event log entries are converted into an object-based model, and then associative rules are extracted dynamically from this model. Secondly, The quality of these rules is improved by filtering insignificant rules. This helps in forming and validating the sequence of events. For forensics investigation, these events can be further examined by a forensics expert. The examination can also be automated by using an action-based domain model. The paper highlights that vulnerability assessment and security configuration activities heavily depend on expert knowledge. Incorporating ML, especially XAI models in this scenario, could aid in automating certain aspects, making the expertise more accessible to non-professionals.

Aditya et al. \citep{aditya2018enabling} present an ATF (Adversary Testing Framework) to test the security robustness of black box DNN (Deep Neural Networks) models in the context of CF. The purpose of developing the ATF is to develop trust in a DNN model. The ATF probes a DNN model's defenses and tries to figure out the workings of the model by calculating the influence of the features on the prediction. The Integrated Gradient \citep{sundararajan2017axiomatic} technique and a surrogate model are used for this purpose. This helps the framework in crafting the inputs to the DNN model for testing its security robustness.

Afzaliseresht et al. \citep{afzaliseresht2019explainable} proposed an XAI model for analyzing security event logs, which can be used during forensics investigation of security events. An apriori algorithm \citep{agrawal1994fast} is used to mine temporal patterns to automatically discover sequential events from a log file, a common application of ML. Storytelling \citep{wu2013internet} is used to present this sequence of events as a knowledge model. This knowledge model highlights the semantic and inferred information from log files in a human-readable format. This provides a contextualized interpretation of security events which can help explain the events and hence reduces the efforts of humans to interpret events.

Costantini et al. \citep{costantini2020assessing} propose a formal framework for assessing the information quality of IoT devices. During forensic analysis of IoT devices using XAI, this framework provides more transparency to the reasoning process during evidence collection for investigation purposes. The authors develop a three-part analytical process for assessing the information quality. Firstly, the preservation of the integrity of collected information. Secondly, the trustworthiness of events is verified by the sources of pieces of evidence. Thirdly, the quality of the discussion about pieces of evidence depends on the competencies of the agents, argumentation abilities, etc. The paper highlights challenges in handling a vast number and extreme variety of IoT items, often lacking physical interfaces, pointing to the need for advanced computational methods. In this work, ML is applied, especially within the XAI-CF framework, which has contributed to handling the complexity and diversity of IoT data, and provided tools for automated analysis and interpretation.

Bolle et al. \citep{bolle2020role} presents a framework based on scientific interpretation principles to evaluate AI forensics analysis systems. This helps in building trust in such systems. The framework comprises three levels/steps, performance evaluation, understandability and transparency evaluation, and forensic evaluation. For the final forensic conclusion to be reliable and trustworthy each of these steps should be formalized. The authors also discussed challenges of measuring understandability, such as subjectivity, data protection, and the risk of making a system simpler to understand but missing an important aspect. The mention of undetected errors or bias resulting in wrong decisions in the paper \citep{bolle2020role} emphasizes the need for transparency and interpretability in automated systems. In the context of forensic analysis, especially when utilizing ML, these techniques become crucial for ensuring that the decision-making process is understandable and reliable. Also, the reference to decisions violating fundamental human rights due to outputs that are not well understood highlights the ethical implications of automated systems. XAI-CF focuses on making AI systems interpretable and explainable, addressing concerns related to accountability and human rights.

Mahajan et al. \citep{mahajan2021explainable} evaluate different AI systems for toxic comment classification. During forensic analysis of social media content, there is a need to check the toxicity of the communication on social media platforms. During experiments, the authors noticed that the standard evaluation metrics, such as accuracy, precision, recall, etc, can be deceiving in evaluating AI systems. They instead use LIME, one of the XAI techniques for model interpretability to evaluate AI models. The results concluded that XAI techniques like LIME are important in selecting the best model and also increasing the user's trust in the model. The paper underscores the significance of model interpretability techniques, specifically mentioning LIME, to understand the decisions made by ML models. This aligns with the core principles of XAI-CF, which emphasizes the interpretability of AI models, especially in sensitive domains like CF.

Nguyen et al. \citep{nguyen2021human} propose to use a human-in-the-loop (an expert forensic analyst) in an XAI system so that both of them complement each other to produce effective and efficient results. The forensic analyst compares the pieces of evidence and provides adjustments to the errors identified in the previous iteration. The input to the XAI model is modified according to these adjustments. This way with each iteration the system improves. In the end, the forensic analyst produces the final report with all the observable pieces of evidence. The proposed framework emphasizes the integration of security analysts or forensic investigators into the man-machine loop. The model is still under development and there are no experimental studies presented in the paper.

Pethe et al. \citep{pethe2022atle2fc} propose a model for explainable IoT forensics using ensemble learning. The use of an ensemble layer comprising various machine-learning models for classification demonstrates the incorporation of ML principles. Ensemble learning involves the combination of multiple models to enhance predictive performance, highlighting the relevance of the ML class. The ensemble layer consists of different ML models for producing the classification. The classification results from the ensemble layer are input to the explainable layer that consists of FPGrowth (Frequent Pattern Growth) and a deep learning classifier. FPGrowth is a recommender that assists in finding mutually dependent events. A deep learning model then categorizes the severity level of these events. The severity level in turn is used to fine-tune the classification and explainability models.

Hall et al. \citep{hall2022proof} evaluate different AI models using LIME. These models were trained to predict image and file content and file system metadata on a virtual hard disk image. These predictions were processed using LIME, an XAI tool that enhances the performance and investigative capabilities of AI's model to assist CF professionals. The authors collected a sample of 23 virtual hard disk files containing a Windows system with a New Technology File System (NTFS) and a single user account. Different file types were created for forensics analysis. After classification, the results were input to LIME for explanation. The results indicate that LIME most of the time was able to explain the classification results but sometimes it failed, such as when feature interaction was involved. The use of LIME to evaluate different AI models for forensics analysis reinforces the connection of this research with XAI-CF.

Jayakumar et al. \citep{jayakumar2022visually} propose a method that enhances the interpretability of deepfake detection models. This method is employed to visually explain why a deepfake detection model classifies a video as a deepfake. The deep fake detection model could play a crucial role in the decision-making process of juries in digital forensic investigations. The authors use Anchors \citep{ribeiro2018anchors} one of the XAI methods, emphasizing the integration of XAI techniques. This shows that XAI methods play a crucial role in providing interpretability and transparency to ML models, especially in contexts where justifications behind tool decisions are essential.

De et al. \citep{de2022interpretable} take a step towards explainable CF by proposing an interpretable federated transformer log learning model for threat detection in a cloud. The authors propose a model for threat detection in a cloud environment. This model is based on a local transformer-based threat detection model trained at each client within an organizational unit. Firstly, a local model is trained for each client using the local dataset of system logs. Secondly, the learned parameters from all the local models are fed to the federated learning server to generate a global federated learning model, which is shared with each client. This cycle is repeated many times to improve threat detection. Federated interpretability weights are computed by aggregating the local interpretability weights of each client. The interpretability module of each client highlights differences in normal and threat sequences that in turn help in the decision-making process. Additionally, it provides forensic investigators with actionable information to effectively analyze system activity and operation.

Bouter et al. \citep{bouter2023protoexplorer} present a system named Protoexplorer for visualizing and interpreting predictions of deepfake video data for forensics analysis. The paper mentions the importance of interpretable ML models in high-stakes settings, especially with the prevalence of complex deep-learning models. Also, the paper introduces prototype-based methods as a promising approach to make deep learning interpretable. Protoexplorer improves the explainability of deepfake video detection models by replacing the prototypes with alternative suggested prototypes. Protoexplorer provides a forensics analyst with meaningful visualizations and intuitive interactions with models. This helps the forensics analyst thoroughly evaluate and explain the models. The main action a forensics analyst can perform is zooming in on a part of the image and observing the contribution of the model to the prediction score. This step is called forensic deepfake analysis where a decision is made if the video is manipulated or not. This decision should be interpreted so that it can be presented in a court of law as a piece of trustworthy evidence. For interpretation, the forensics analyst explores and refines the model and provides an explanation and understanding of how the model aided in making the decision.

\hl{Alam et al.{~\citep{alam2024sift}} present SIFT, a machine-learning–based system for automated file-type identification designed for digital forensic analysis. Rather than depending only on traditional signature or header inspection, SIFT employs AI-driven classification to determine file types even when metadata is missing, manipulated, or deliberately concealed. A major contribution is the use of XAI techniques to make the system’s outputs more transparent and defensible in forensic settings. Their approach involves training deep learning models on byte-level features and applying explanation tools such as LIME and SHAP to highlight which regions or attributes influenced each prediction.
Khalid et al.{~\citep{khalid2024towards}} introduce a comprehensive framework that integrates XAI across the CF pipeline, from evidence acquisition and feature extraction to model inference, explanation generation, and presentation to investigators or courts. The aim is to standardize processes and enhance the legal and operational defensibility of AI-generated results.

Hermosill et al.{~\citep{hermosilla2025explainable}} analyze SHAP and LIME explanations applied to two classifiers, XGBoost and TabNet, for network intrusion detection using the UNSW-NB15 dataset. Their evaluation covers explanation stability, global interpretability, and trade-offs between the two post-hoc methods. They also compare classifier performance, noting that XGBoost achieved higher validation accuracy than TabNet.
Khalid et al.{~\citep{khalid2025bridging}} investigate combining unsupervised learning with XAI approaches for forensic scenarios characterized by massive unlabeled datasets (e.g., memory forensics, large disk images). Their work emphasizes clustering and anomaly detection augmented with feature-level and prototype-based explanations to reveal novel artifacts and assist investigators in prioritizing analysis.}

\subsubsection{Summary}

\hl{This compilation of research emphasizes the swiftly evolving role of XAI across diverse CF tasks. These tasks encompass log analysis, intrusion detection, deepfake investigations, file-type identification, IoT forensics, and federated systems. The key findings from these studies can be categorized into six primary themes, highlighting the notable impact and advancements that XAI brings to the field of CF.}

\begin{enumerate}
    \item 
    \hl{XAI is rapidly becoming a foundational requirement for cyber forensics, driven by legal, ethical, and operational constraints.}
    \item 
    \hl{AI paired with XAI significantly improves forensic workflows, from log analysis and file-type identification to deepfake detection and IoT event reconstruction.}
    \item 
    \hl{Explanations, not just predictions, are crucial for reliability, validation, and courtroom defensibility.}
    \item 
    \hl{Unsupervised and adversarial settings benefit especially from XAI, where explanations reveal hidden patterns, anomalies, and system vulnerabilities.}
    \item 
    \hl{A shift is underway from isolated XAI tools to integrated, pipeline-level frameworks that support end-to-end forensic analysis.}
    \item 
    \hl{Human expertise remains essential, with XAI functioning as an augmentation tool rather than a replacement.}
\end{enumerate}


\section{Challenges and Future Works}\label{sec:challenges}

\hl{In this section, we discuss some of the challenges that need to be resolved for the promising development and use of a practical XAI-CF system. We also provide and recommend solutions to these challenges as part of future work. Finally, we introduce a five-layer framework for an XAI-CF system, while leaving the detailed ontology development for future work.}

\subsection{Human Computer Interaction (HCI)}

Interaction with an XAI-CF system allows the user to ask follow-up questions/queries to better understand and evaluate the explanations. Another major reason for interacting with an XAI-CF system is to include a human-in-the-loop that can produce valuable, practical, and structured explanations. To build such systems that are successful in practice, we need to provide the user with a usable interactive XAI-CF system. Human-Computer Interaction (HCI) and XAI are upcoming research areas and very limited research \citep{wolf2019explainability,shneiderman2020bridging} has been reported in this area, and very few \citep{bouter2023protoexplorer} on HCI and XAI-CF. As part of the CF research community, this is our responsibility to build the new XAI-CF systems interactive from the ground up. There are already several research works on interacting with AI systems \citep{kulesza2012tell, kulesza2009fixing, lee2017human, stumpf2009interacting, yang2013learning}.
Recently developed ChatGPT \citep{ray2023chatgpt, fui2023generative}, the new AI language model has revolutionized the approach in AI to human-model-interaction. ChatGPT is being used in various domains, such as business, education, and healthcare etc. The potentials of ChatGPT are unlimited, but it can be a double-edged sword \citep{fui2023generative}. For example, for businesses, it can create creative content such as ideas for advertisements, and on the other hand, it can create hallucinations and produce fake information. One of the future works is to explore the use of ChatGPT to improve HCI in XAI-CF systems and identify fake or false information for its successful use. Consideration of ethical standards and security measures should be paramount in the development process of interactive XAI-CF systems.
\hl{Francisco Herrera{~\citep{herrera2025reflections}} argues that the evolution of XAI is not just a technical problem (making black‐box models interpretable), but fundamentally a sociotechnical challenge. From an HCI standpoint, it’s not enough that a model can explain itself; these explanations must be meaningfully interpretable and actionable by human collaborators.

A practical XAI-CF system is human-centered, because without human intervention and understanding CF produces no results. Therefore, HCI for such a system should be guided by human values and needs. One such challenge is socio-technical and is about interacting with an XAI-CF system by a person (e.g., law enforcement community) who has little knowledge of such systems. For example, retraining a model on new data is an advanced skill and requires setting up hyperparameters that influence the learning process. One solution is to transfer such knowledge to the law enforcement community or develop new HCI tools that facilitate and automate such actions with ease and without requiring advanced skills or knowledge. We recommend the latter option, which makes the solution general and is human-centered.}

\subsection{Bias Management}

\hl{In an XAI-CF system, the pieces of evidence produced should be unbiased. A bias can occur in an XAI-CF system for different reasons: Biased training data, such as a manually labeled dataset by a forensics analyst, that can affect the credibility of the model's output; Human-infused bias when interacting with XAI-CF; Stale data, such as when out-of-date forensics datasets are used for training. Interested readers are referred to} \citep{mehrabi2021survey, schwartz2022towards} \hl{for a detailed list and discussion of AI biases. The problem of bias in AI is neither new nor unique, and it is not possible to eliminate bias from an AI system. It is an ongoing process, and there are many approaches for mitigating AI bias, such as training workshops that can employ procedures for unbiasing humans that interact with AI and the development of more diverse and comprehensive training data that addresses biases.}

\hl{Hanna et al.{~\citep{hanna2025ethical}} provide a foundational overview of the major ethical challenges associated with AI systems, with a strong emphasis on algorithmic bias and its implications across the AI lifecycle. Roselli et al.{~\citep{roselli2019managing}} recommend a set of processes to mitigate the impact of bias in AI systems. Schwartz et al.{~\citep{schwartz2022towards}} describe three major challenges of mitigating AI bias: datasets, testing, and human factors. The authors also provide recommendations to address these challenges. Stack et al.{~\citep{slack2020fooling}} provided a fooling method that created highly biased classifiers and was able to deceive state-of-the-art XAI techniques, such as LIME and SHAP, in producing simple explanations that did not reflect the underlying biases in the system. There is a need to improve these state-of-the-art XAI techniques and, in parallel, develop new methods and approaches that can successfully manage the AI biases. Legal protection offered by non-discrimination laws is challenged when AI, not humans, discriminate{~\citep{wachter2021fairness}}. There is also a need to devise laws, regulations, and standards that not only address human but also machine biases. We should also establish a continuous monitoring mechanism within the XAI-CF system to detect and address biases in real-time.}

\subsection{Adversarial Attacks}

\hl{XAI-CF can help an adversary without much effort to learn about AI forensic methods and plan new attack surfaces (anti-forensics) in a much better way. Is it possible, and how to evade this situation? Adversarial attacks are one of the major challenges of an AI system}~\citep{du2020sok,chakraborty2021survey}. \hl{The adversary can manipulate the input to force the XAI-CF system to produce incorrect output. A pre-trained model used during an investigation loses its effectiveness, and hence, the pieces of evidence obtained lose their credibility. Different techniques are used by an adversary to attack an XAI system, such as data poisoning} \citep{baniecki2022fooling,aivodji2022fooling,baniecki2022manipulating}, \hl{adversarial example{~\citep{huang2023safari}}, and trust manipulation {\citep{lakkaraju2020fool}}, etc. A detailed list of attacks and their defenses is listed and discussed in}~\citep{baniecki2023adversarial}. \hl{The surveys provided in}~\citep{baniecki2023adversarial, vadillo2025adversarial} \hl{reiterate about the apparent insecurities in XAI, that there are still several unaddressed attacks on explanation methods.}

To prepare and train an XAI-CF system against adversarial attacks, one should know about the attack. Adversarial ML \citep{huang2011adversarial} is a technique that learns from previous attacks by adversaries and plans new attacks that deteriorate the performance of the classifier. To counter these attacks, different approaches are discussed in \citep{martins2020adversarial} to adapt the classifiers to improve their classification. Generative Adversarial Networks \citep{goodfellow2020generative} is another technique being applied to give insights into specific behaviors of adversaries to improve the defense against their attacks. The two techniques, Adversarial ML and Generative Adversarial Networks, can be used to improve the defenses against adversarial attacks. An adversarial attack can also be used as an anti-forensics technique \citep{garfinkel2007anti}. There are several research works carried out to detect anti-forensics activities~\citep{garfinkel2007anti, rekhis2011system, bhatt2017machine, hoelz2009artificial, mitchell2010use, sun2018novel, yu2016multi, chen2018densely, abozaid2023deployment, goel2023approach}. To prevent anti-forensics activities, only one solution is provided, and that is to preserve the privacy of the forensics evidence~\citep{armknecht2015privacy, nieto2016digital, nieto2017methodology}, including digital files, emails, log files, and other documents. \hl{To develop counter-anti-forensics techniques is an open area of research}~\citep{du2020sok, surakanti2025countering}.

\subsection{Oversimplification}

\hl{The more accurate an AI system is, the more difficult it is to explain and vice versa. One of the solutions to make AI explainable is to simplify the AI system. But there is a danger lurking in this process; we oversimplify the AI system to the extent that it may not produce the right results. Moreover, simplifying the explanation of a complex AI system, such as neural networks, will inevitably leave out important details that may hinder the discovery of new knowledge, i.e., the learning part of an AI system. To mitigate such risks, we have to develop techniques that exchange accuracy with explainability in a controlled manner or provide explanations without affecting the accuracy. A lot of work has been done in the area of explainability in cognitive psychology} \citep{von2004explanation, keil2006explanation, schank2013explanation}, \hl{but the same has not been systematically transferred to AI. Miller et al.} \citep{miller2019explanation} \hl{discuss in detail how we should develop methods and approaches to infuse this valuable research into XAI. Humans select explanations that are simpler than their full counterparts} \citep{miller2019explanation}. \hl{Explanation of the XAI-CF system should be provided and simplified in a way that is to the satisfaction of the stakeholders of CF and also does not compromise the accuracy of the system. Thagard {\citep{thagard1989explanatory}} gave seven foundational principles that explanation must comply with to be acceptable. Out of these seven, he contends that the dependency of a better explanation on two principles is simpler and more general. If all other things are equal, \textit{simpler} explanations -- those that cite fewer causes, and more \textit{general} explanations -- that explain more events are better. Later, this theory, that humans prefer simpler and more general explanations, is tested and proven true through an empirical study} \citep{read1993explanatory}.

\hl{The key lies in shifting explainability from static, post-hoc visualizations toward dynamic, user-driven interaction, where users can probe, query, and manipulate model explanations in real time{~\citep{speckmann2025ixaii}}. In this context and as mentioned before, we recommend a simpler interactive visualization (e.g., a causal loop or stack and flow diagram) depicting more events that can provide a better explanation of an XAI-CF system.}

\subsection{CF and AI Chasm}

The disconnect between CF and AI communities \citep{baggili2019founding} constitutes one of the major socio-technical challenges to developing a practical XAI-CF system. There is a lack of collaboration between these two communities that is not letting researchers from both these areas work together to bridge this gap. \hl{Moreover, the majority of the CF practitioners have limited knowledge of AI}~\citep{sanchez2019practitioner, abraham2021automatically, ibrahim2025artificial}. \hl{For example, as mentioned before, one solution to improve HCI in an XAI-CF system is to transfer basic AI knowledge to CF personnel. Furthermore, including a human-in-the-loop{~\citep{amaliah2025human}} (CF practitioner) complements an XAI-CF system. This is only possible if there is a strong collaboration between these two communities, and CF practitioners and other stakeholders start learning the basics of AI.}

Currently, researchers have pointed out this lack of disconnect~\citep{baggili2019founding, sanchez2019practitioner, abraham2021automatically} between CF and AI, but there is no concrete work done to bridge this gap. Firstly, we recommend conducting workshops that bring together researchers and practitioners from both CF and AI communities to discuss and raise awareness of this topic. We also need to include law enforcement agencies in such workshops. Secondly, we need to develop standards, notations, formalisms of concepts, and explanations that are easy to understand by both communities. In this context, for easier transfer of results and information, there is a need to use systematic definitions and simpler and more general explanations of XAI systems. This way it will be easier for a CF practitioner to learn and interact with an XAI-CF system.

\subsection{Towards an XAI-CF System: A Conceptual Framework}

\begin{figure*}[!ht]
	\centering
	\includegraphics[scale=0.42]{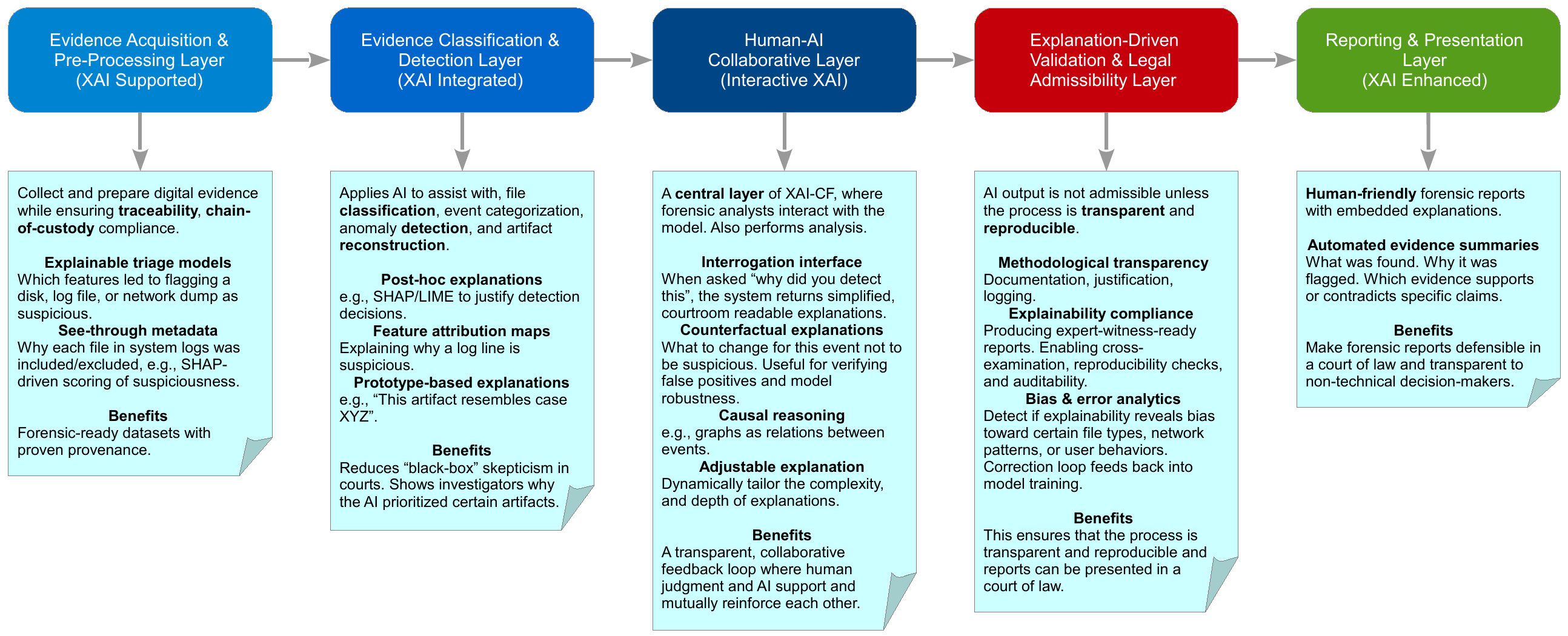}
	\caption{\hl{Towards an XAI-CF system: A layered architecture.}}
	\label{fig:proposed_framework}
\end{figure*}

\hl{Existing studies rarely integrate explainability, legal defensibility, and human-AI collaboration in a unified architecture. To bridge this gap, we propose a five-layer framework that embeds XAI across the entire forensic lifecycle. Figure{~\ref{fig:proposed_framework}} provides a high-level overview of the proposed conceptual framework. The framework employs a layered architecture consisting of five interconnected layers, with each layer relying on the output of the one before it.}

\hl{The proposed XAI-CF framework uniquely integrates XAI principles throughout the entire digital forensic lifecycle. Unlike existing works that apply AI primarily for detection or classification, this framework:}

\begin{itemize}
    \item 
    \hl{Enhances transparency and interpretability of all AI-driven steps.}
    \item 
    \hl{Supports human–AI collaboration during investigation.}
    \item 
    \hl{Aligns forensic outputs with legal admissibility requirements.}
\end{itemize}

\hl{As such, it provides the first holistic blueprint for embedding explainability into modern CF practice, while leaving the detailed ontology for future work.}

\section{Threats to Validity}\label{sec:threatstovalidity}

For conducting and completing an unbiased study and literature review, and to extract the information correctly, we mostly used three reviewers for collecting the data. Because of the scarcity of resources, we used the same three reviewers to review all the papers. There is no unified state-of-the-art approach or technique for comparing our work with others; therefore, our comparison is only indicative and symbolic. To make the comparison and analysis of the research works presented in this paper as fair as possible and to mitigate the threats mentioned above, we developed and used the research methodology as discussed in section \ref{sec:rm}. Further, we used the principle of triangulation \citep{yin2015qualitative,patton2014qualitative} to increase the credibility of the study carried out in this paper. The triangulation principle is the way of seeking at least three ways of verifying a procedure, piece of data, or finding. We used scite, an AI-powered research engine, Google Scholar search, and manual confirmation of the collected data from different sources, i.e., \textit{data triangulation}. We used three reviewers to check and confirm the procedures and methods adapted for the study carried out in this paper, i.e., \textit{methodological triangulation}. Besides all these efforts for objectivity, there may be chances of relativism \citep{yin2015qualitative,def-relativism} present in the study carried out in this paper.

\section{Conclusion}\label{sec:conclusion}

CF is facing many new challenges, such as the rise of complex cyber devices, the proliferation of operating systems and file formats, pervasive encryption, the use of the cloud for remote processing and storage, and legal standards. Overcoming these challenges requires new techniques, such as those from the field of AI. To apply these techniques successfully in CF, we need to justify and explain the results to the stakeholders of CF, such as forensic analysts and members of the court, for them to make an informed decision. If we want to apply AI successfully in CF, there is a need to develop trustworthy, authentic, interpretable, understandable, and interactive AI systems. An XAI-CF system can play this role, but it is indispensable and is still in its infancy. In this paper, we explore and make a case for the significance and advantages of XAI-CF. We strongly emphasize the need to build a successful and practical XAI-CF system and discuss some of the main requirements and prerequisites of such a system. We present a formal definition of the terms CF and XAI-CF and a comprehensive literature review of previous works that apply and utilize XAI to build and increase trust in CF. To make the reader familiar with the study carried out in this paper, in addition to the background, we also present a critical and short review of the works carried out in the last ten years in XAI and CF. We discuss some challenges facing XAI-CF, such as adversarial attacks, bias management, oversimplification, the CF and AI chasm, and human-computer interaction. We also provide some concrete solutions to these challenges. We identify key insights and future research directions for building XAI applications for CF. \hl{We introduce the first XAI-CF framework, designed to address the long-standing challenges of opacity, limited interpretability, and poor auditability in AI-assisted forensic workflows.} This paper is an effort to explore and familiarize the readers with the role of XAI applications in CF, and we believe that our work provides a promising basis for future researchers interested in XAI-CF.


\end{document}